\documentclass[prd,a4,12pt,showpacs]{revtex4}

\usepackage{amssymb}
\usepackage{amsmath}
\usepackage{graphicx}
\usepackage{color}

\begin{document}

\title{Supersymmetric QCD corrections to single top quark production
at hadron colliders}
\author{Jia Jun Zhang, Chong Sheng Li, Zhao Li, and Li Lin Yang\\
Department of Physics, Peking University, Beijing 100871, China}
\date{\today}

\begin{abstract}
We present the calculations of the supersymmetric QCD corrections to
the total cross sections for single top production at the Fermilab
Tevatron and the CERN Large Hadron Collider in the minimal
supersymmetric standard model. Our results show that for the
s-channel and t-channel, the supersymmetric QCD corrections are at
most about $1\%$, but for the associated production process
$pp\rightarrow tW$, the supersymmetric QCD corrections increase the
total cross sections significantly, which can reach about $6\%$ for
most values of the parameters, and the supersymmetric QCD
corrections should be taken into consideration in the future high
precision experimental analysis for top physics.
\end{abstract}

 \maketitle

\newpage
\section{Introduction}

The search for single top quark production is one of the major aims
of both the Fermilab Tevatron and the CERN Large Hadron
Collider(LHC)  \cite{topquark,wagner}, because it can probe the
electroweak sector of the Standard Model(SM), in contrast with the
dominant QCD pair production mechanism, and provide a consistency
check on the measured parameters of the top quark in the QCD pair
production process
 \cite{topquark123,topquark129,topquark130,topquark131}. Further more, the
mass of top quark is comparable to the electroweak(EW) symmetry
breaking scale, and it can play a role of wonderful probe for the EW
symmetry breaking mechanism and new physics, especially the minimal
supersymmetry standard model(MSSM) as a very popular model beyond
the SM, via non-standard couplings \cite{topquark129,
topquark132,topquark133,topquark134,topquark135,topquark142,liu1,liu2},
loop effects \cite{topquark136,topquark137,topquark138,topquark139,
topquark140}, etc., in the processes for single top quark
production.

At the LHC single top quarks are produced primarily via the
t-channel \cite{topquark120},
$$q+b\rightarrow q'+t,$$
the quark annihilation process(s-channel)
\cite{topquark121,topquark122},
$$q+\bar{q'}\rightarrow t+\bar{b},$$
and the associated production process
\cite{topquark123,topquark124},
$$g+b\rightarrow t+W^-,$$
which can reliably be predicted in the SM, and their leading
order(LO) results are summarized in Table \ref{treelevelresult}
\cite{wagner}.
\begin{table}[!ht]
\caption{The LO results for single top production at the LHC}
\label{treelevelresult}
\begin{center}
\begin{tabular}{|c|c|c|c|}\hline
process:&t-channel&s-channel&Wt \\\hline
$\sigma(pb)$:&$156\pm8$&$6.6\pm0.6$&$14.0^{+3.8}_{-2.8}$\\\hline
\end{tabular}
\end{center}
\end{table}
As a high-luminosity LHC would run soon and allow accurate
measurements of these cross sections with a statistical uncertainty
of less than $2\%$
\cite{topquark,topquark124,topquark149,topquark152,topquark30}, at
this level of experimental accuracy, calculations of the radiative
corrections are necessary to test the predictions of the SM and to
investigate loop effects arising from new physics. The QCD
corrections to the total cross sections of the three channels are
about $-10\%$, $+50\%$ and $+50\%$, respectively
\cite{wagner,topquark124,topquark125,topquark126,myself2,softgluon}.
For t-channel, the SM EW corrections are about $10\%$ and the
supersymmetric(SUSY) EW corrections are a few percent \cite{myself1}
at the LHC, for s-channel, the combined effects of SUSY QCD, SUSY EW
and the Yukawa couplings can exceed $10\%$ for small $\tan\beta(<2)$
but are only a few percent for $\tan\beta>2$ at the Tevatron
\cite{topquark138, topquark139}, and for the associated production
at the LHC, the SM EW and the SUSY corrections have been calculated
and analyzed in Ref.\cite{plusgbew}. But at the LHC, the SUSY QCD
corrections to the three channels have not been calculated yet. As
the associated production process involves the QCD coupling in
initial state, obviously, the SUSY QCD corrections are significant
for this channel. In order to fill in the blanks in the relevant
radiative corrections to the channels considered here, in this paper
we present the calculations of the SUSY QCD corrections to the three
channels at the CERN LHC, mainly concerning the associated
production channel, and also give the updated numerical calculations
to the s-channel process at the Tevatron\cite{li}.

Our paper is organized as follows. In Section II we will give the
analytic results in terms of the well-known standard notation of
one-loop Feynman integrals for the associated channel, s-channel and
t-channel. In Section III we will present our numerical results with
discussions of their implications.

\section{Analytic Results}

\subsection{Associated Production}

In our calculations we use dimensional reduction to control all the
ultraviolet divergences in the virtual loop corrections. We adopt
the on-mass-shell renormalization scheme\cite{li11} for the top
quark mass and the wave functions renormalization, and set all the
other quark masses as zero. The QCD coupling constant $g_s$ is
renormalized in the modified minimal subtraction
scheme($\overline{\mathrm{MS}}$) except that the divergences
associated with the top quark and colored SUSY particle loops are
subtracted at zero momentum\cite{myself6}. Denoting $\psi_{q0}$,
$m_{q0}$, $g_{s0}$ and $A_{\mu0}$ as bare quark wave functions,
quark masses, strong coupling constant and gluon wave function
respectively, the relevant renormalization constants are then
defined as
\begin{eqnarray}
\psi_{q0}&=&(1+\delta Z_L^{qq})^{1/2}\psi_{qL}+(1+\delta
Z_{R}^{qq})^{1/2} \psi_{qR},\notag\\
m_{q0}&=&m_q+\delta m_q,\notag
\end{eqnarray}
\begin{eqnarray}
A_{\mu0}&=&(1+\delta Z_{AA})^{1/2}A_{\mu},\notag\\
g_{s0}&=&(1+\delta Z_g)g_s,
\end{eqnarray}

\begin{figure}
\begin{center}
\scalebox{0.6}{\includegraphics{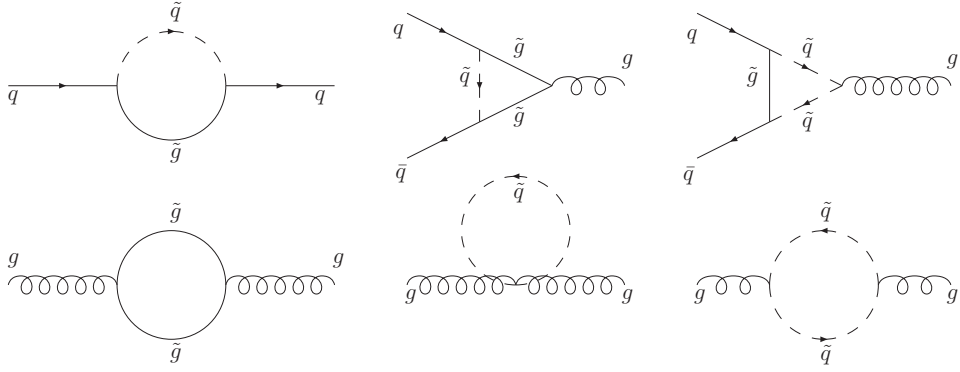}}
\end{center}
\caption{The self-energy and vertex diagrams for calculating the
renormalization constants.} \label{renconst}
\end{figure}

After calculating the self-energy and vertex diagrams in
Fig.\ref{renconst}, we obtain the explicit expressions of all
necessary renormalization constants as follows:
\begin{eqnarray}\label{RC}
\delta
Z_g&=&-\frac{\alpha_s(\mu^2)}{4\pi}\{\frac{\beta_0}{2}[\frac{1}{\epsilon}-\gamma_E+\ln(4\pi)]+
\frac{N}{3}\ln(\frac{M_{\tilde{g}}^2}{\mu^2})+
\sum_{u=u,c,t}^{i=1,2}\frac{1}{12}\ln(\frac{M_{\tilde{u}_i}^2}{\mu^2})+
\sum_{d=d,s,b}^{i=1,2}\frac{1}{12}\ln(\frac{M_{\tilde{d_i}}^2}{\mu^2})\}
,\notag\\
\delta Z_{AA}&=&-\frac{3\alpha_s}{2\pi}B_1(0, M_{\tilde{g}}^2,
M_{\tilde{g}}^2)+ \frac{3\alpha_s}{\pi}\frac{\partial}{\partial p^2}
B_{00}(0, M_{\tilde{g}}^2, M_{\tilde{g}}^2)\notag\\
&&-\frac{\alpha_s}{2\pi}\sum_{u}^{3}\sum_{i}^{2}
\frac{\partial}{\partial p^2} B_{00}(0, M^2_{\tilde{u}_i},
M^2_{\tilde{u}_i})-\frac{\alpha_s}{2\pi}\sum_{d}^{3}\sum_{i}^{2}
\frac{\partial}{\partial p^2} B_{00}(0, M^2_{\tilde{d_i}},
M^2_{\tilde{d_i}}),\notag\\
\delta
m_t&=&\frac{2\alpha_sM_{\tilde{g}}}{3\pi}\sum_{i}^{2}B_0(m_t^2,
M_{\tilde{g}}^2, M^2_{\tilde{t}_i})U^t_{i1}U^{t*}_{i2}+
\frac{\alpha_sm_t}{3\pi}\sum_{i}^{2}B_1(m_t^2, M_{\tilde{g}}^2,
M^2_{\tilde{t}_i})(U^t_{i1}U^{t*}_{i1} + U^t_{i2}U^{t*}_{i2}),\notag\\
\delta
Z^{tt}_{(L,R)}&=&-\frac{2\alpha_s}{3\pi}\sum_{i}^{2}B_1(m_t^2,
M_{\tilde{g}}^2, M_{\tilde{t}_i}^2)U^t_{i(1,2)}U^{t*}_{i(1,2)}-
\frac{4\alpha_sM_{\tilde{g}}m_t}{3\pi}\sum_i^{2}\frac{\partial}
{\partial p^2}B_0(m_t^2, M_{\tilde{g}}^2, M^2_{\tilde{t}_i})
U^t_{i1}U^{t*}_{i2}\notag\\
&& - \frac{2\alpha_sm_t^2}{3\pi}\sum_i^2\frac{\partial}{\partial
p^2} B_1(m_t^2, M_{\tilde{g}}^2,
M^2_{\tilde{t}_i})(U^t_{i1}U^{t*}_{i1}+ U^t_{i2}U^{t*}_{i2}),\notag\\
\delta Z^{qq}_{(L,R)}&=&-\frac{2\alpha_s}{3\pi}\sum_i^2B_1(0,
M_{\tilde{g}}^2, M_{\tilde{q}_i}^2)U^q_{i(1,2)}U^{q*}_{i(1,2)}\qquad
q=u,d,s,c,b,
\end{eqnarray}
where $\beta_0=[-\frac{2}{3}(N+1)-\frac{1}{3}(n_f+1)]$,
$B_{ijk\cdots}$ are the two-point functions, $U^q_{ij}$ are the
mixing matrices of the squarks, $\gamma_E$ is the Euler constant,
and $\mu$ is the renormalization scale.

Including the SUSY QCD corrections, the renormalized amplitudes can
be written as
\begin{eqnarray}
M_{ren}^A&=&M_0^A+M_{vir}^A+M_{count}^A,
\end{eqnarray}
where $M_0^A$ is the LO amplitude, $M_{vir}^A$ represents the SUSY
QCD corrected amplitude from the one-loop self-energy, vertex, and
box diagrams, and $M_{count}^A$ is the corresponding counterterm for
the self-energy corrections and vertex corrections, respectively.

\begin{figure}
\begin{center}
\scalebox{0.6}{\includegraphics[340,500][440,700]{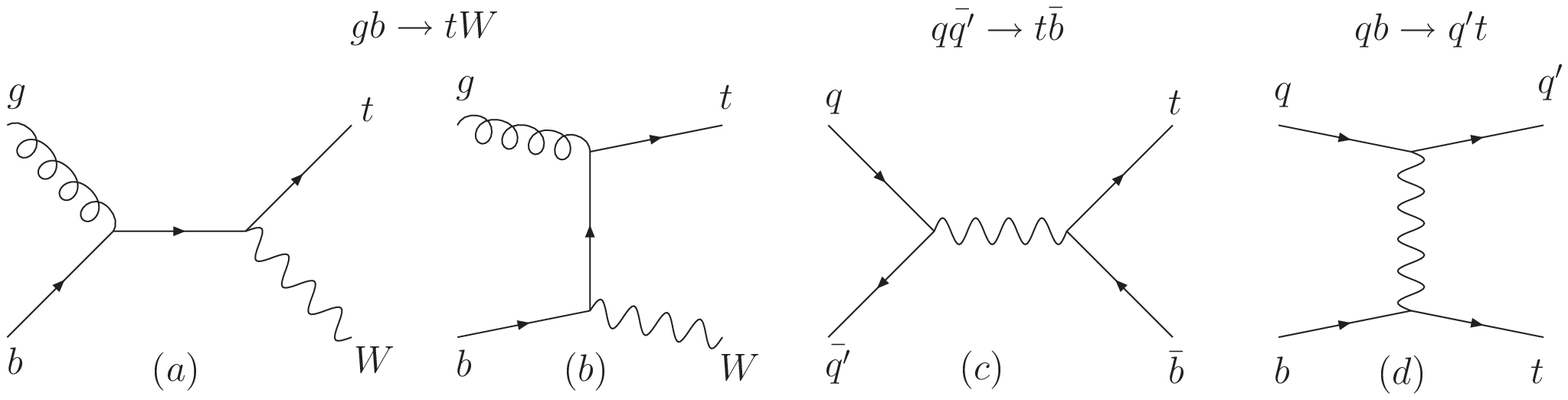}}
\end{center}
\caption{Tree level Feynman diagrams for associated production.}
\label{tree}
\end{figure}

The related Feynman diagrams which contribute to the LO amplitude
$M_0^A$ are shown in Fig.\ref{tree} (a) and (b). The LO amplitude
$M_0^A$ is given by
\begin{eqnarray}
M_0^A&=&M_s^A+M_t^A\notag\\
&=&\frac{eg_sV_{tb}}{\sqrt{2}\sin\theta_W}[\frac{1}{s}(2A_{10}-A_7
-2A_8)+\frac{1}{t-m_t^2}(2A_{15}-2A_9-A_7)],
\end{eqnarray}
where $V_{ij}$ are the elements of CKM matrix, s, t, u are the
Mandelstam invariants, which are defined as,
\begin{eqnarray}
s&=&(k_1+k_2)^2=(k_3+k_4)^2,\notag\\
t&=&(k_1-k_3)^2=(k_2-k_4)^2,\notag\\
u&=&(k_1-k_4)^2=(k_2-k_3)^2,
\end{eqnarray}
where $k_1$ and $k_2$ denote the momentum of the incoming particles,
$k_3$ and $k_4$ the outgoing particles, while $A_m$ are the reduced
standard matrix elements given by
\begin{eqnarray}\label{FI}
A_{1,22}&=&(\epsilon^a_1\cdot
\epsilon^*_4)\bar{u}^b(k_3)P_{R,L}u^c(k_2)
(T^a)_{bc},\notag\\
A_{2,3}&=&(\epsilon^a_1\cdot k_2)(\epsilon^*_4\cdot k_{1,2})
\bar{u}^b(k_3)P_Ru^c(k_2)(T^a)_{bc},\notag\\
A_{4,5}&=&(\epsilon^a_1\cdot k_3)(\epsilon^*_4\cdot k_{2,1})
\bar{u}^b(k_3)P_Ru^c(k_2)(T^a)_{bc},\notag\\
A_{6,17}&=&\bar{u}^b(k_3)P_{R,L}\not{\epsilon^a_1}\not{\epsilon^*_4}u^c(k_2)
(T^a)_{bc},\notag\\
A_{7,27}&=&\bar{u}^b(k_3)P_{R,L}\not{\epsilon^a_1}\not{\epsilon^*_4}\not{k_1}u^c(k_2)
(T^a)_{bc},\notag\\
A_{8,9}&=&(\epsilon^a_1\cdot
k_{2,3})\bar{u}^b(k_3)P_R\not{\epsilon^*_4}u^c(k_2)(T^a)_{bc},\notag\\
A_{10,32}&=&(\epsilon^a_1\cdot \epsilon^*_4)
\bar{u}^b(k_3)P_{R,L}\not{k_1}u^c(k_2)(T^a)_{bc},\notag\\
A_{11,12}&=&(\epsilon^*_4\cdot k_2)(\epsilon^a_1\cdot k_{2,3})
\bar{u}^b(k_3)P_R\not{k_1}u^c(k_2)(T^a)_{bc},\notag\\
A_{13,14}&=&(\epsilon^*_4\cdot k_1)(\epsilon^a_1\cdot k_{2,3})
\bar{u}^b(k_3)P_R\not{k_1}u^c(k_2)(T^a)_{bc},\notag\\
A_{15,16}&=&(\epsilon^*_4\cdot k_{1,2})
\bar{u}^b(k_3)P_R\not{\epsilon_1^a}u^c(k_2)(T^a)_{bc},\notag\\
A_{18,19}&=&(\epsilon^*_4\cdot k_{1,2})
\bar{u}^b(k_3)P_R\not{\epsilon_1^a}\not{k_1}u^c(k_2)
(T^a)_{bc},\notag\\
A_{20,21}&=&(\epsilon^a_1\cdot k_2)(\epsilon^*_4\cdot k_{1,2})
\bar{u}^b(k_3)P_Lu^c(k_2)(T^a)_{bc},\notag\\
A_{23,24}&=&(\epsilon^a_1\cdot k_3)(\epsilon^*_4\cdot k_{2,1})
\bar{u}^b(k_3)P_Lu^c(k_2)(T^a)_{bc},\notag\\
A_{25,26}&=&(\epsilon^*_4\cdot k_{1,2})
\bar{u}^b(k_3)P_L\not{\epsilon_1^a}u^c(k_2)(T^a)_{bc},\notag\\
A_{28,29}&=&(\epsilon^*_4\cdot k_{1,2})
\bar{u}^b(k_3)P_L\not{\epsilon_1^a}\not{k_1}u^c(k_2)
(T^a)_{bc},\notag\\
A_{30,31}&=&(\epsilon^a_1\cdot k_{2,3})
\bar{u}^b(k_3)P_L\not{\epsilon^*_4}u^c(k_2)(T^a)_{bc},\notag\\
A_{33,35}&=&(\epsilon^*_4\cdot k_{2,1})(\epsilon^a_1\cdot k_2)
\bar{u}^b(k_3)P_L\not{k_1}u^c(k_2)(T^a)_{bc},\notag\\
A_{34,36}&=&(\epsilon^a_1\cdot k_3)(\epsilon^*_4\cdot k_{2,1})
\bar{u}^b(k_3)P_L\not{k_1}u^c(k_2)(T^a)_{bc},\notag\\
A_{37,38}&=&(\epsilon_1^a\cdot k_{3,2})
\bar{u}^b(k_3)P_L\not{\epsilon^*_4}\not{k_1}u^c(k_2)(T^a)_{bc}.
\end{eqnarray}

The relevant Feynman diagrams of the SUSY QCD corrected amplitude
$M_{vir}^A$ are shown in Fig.\ref{gb2tw}, and $M_{vir}^A$ can be
written as
\begin{eqnarray}
M_{vir}^A&=&M_{self}^A+M_{vertex}^A+M_{box}^A,
\end{eqnarray}
where $M_{self}^A$, $M_{vertex}^A$ and $M_{box}^A$ come from
self-energy diagrams, vertex diagrams and box-diagrams as shown in
Fig.\ref{gb2tw}, respectively. Their explicit expressions are given
by
\begin{eqnarray}\label{general}
&&M_{self}^A=\sum_{m=1}^{38}f_m^{self}A_m,\quad
M_{vertex}^A=\sum_{m=1}^{38}f_m^{v}A_m,\quad
M_{box}^A=\sum_{m=1}^{38}f_m^{b}A_m,
\end{eqnarray}
where $f_m^{self},f_m^{v},f_m^{b}$ are the form factors, which are
given explicitly in Appendix.

The counterterms $M_{count}^A$, the corresponding diagrams of which
are shown in Fig.\ref{counterterm}, can be written as follows
\begin{eqnarray}
M_{count}^A&=&\delta M_{self}^A+\delta M_{vertex}^A,
\end{eqnarray}
with
\begin{eqnarray}
\delta M_{self}^A&=&\delta M_{self}^{1}+\delta M_{self}^{2}\notag\\
&=&\frac{eg_sV_{tb}}{2\sqrt{2}\sin\theta_W}\{\
\frac{A_{17}}{t-m_t^2}[m_t(\delta Z_L^{tt}-\delta Z_R^{tt})+
2\delta m_t]-\frac{2\delta Z_L^{bb}}{s}(A_7-2A_8+2A_{10})\notag\\
&&+\frac{2}{(t-m_t^2)^2}(A_7+2A_9-2A_{15})[(m_t^2-t)\delta
Z_L^{tt}+2m_t\delta m_t]\ \},\\
\delta M_{vertex}^A&=&\sum_{n=1}^4\delta M_{vertex}^{n},\\
\delta M_{vertex}^{1}&=&\frac{eg_sV_{tb}}
{2\sqrt{2}s\sin\theta_W}(A_7+2A_8-2A_{10})(\delta Z_L^{bb}+\delta
Z_L^{tt}),\notag\\
\delta M_{vertex}^{2}&=&\frac{eg_sV_{tb}
}{2\sqrt{2}\sin\theta_W(t-m_t^2)}(A_7+2A_9-2A_{15})(\delta
Z_L^{bb}+\delta Z_L^{tt}),\notag\\
\delta
M_{vertex}^{3}&=&\frac{eg_sV_{tb}}{2\sqrt{2}\sin\theta_W(t-m_t^2)}[
2m_tA_{17}(\delta Z_R^{tt}-\delta Z_L^{tt})\notag\\
&&\qquad\qquad
+(A_7+2A_9-2A_{15})(2\delta Z_g+2\delta Z_L^{tt}+\delta Z_{AA})],\notag\\
\delta
M_{vertex}^{4}&=&\frac{eg_sV_{tb}}{2\sqrt{2}s\sin\theta_W}(2\delta
Z_g + \delta Z_{AA} + 2\delta Z_L^{bb})(A_7 + 2A_8 - 2A_{10}).
\end{eqnarray}

The partonic cross section can be written as
\begin{eqnarray}
\hat{\sigma}&=&\int_{-1}^1\mathrm{d}z\frac{1}{32\pi
s^2}\lambda^{1/2}\overline{
|M_{ren}^A|^2}=\int_{t_-}^{t_+}\mathrm{d}t\frac{1}{16\pi
s^2}\overline{ |M_{ren}^A|^2},
\end{eqnarray}
where $\lambda\equiv(s-m_t^2+m_W^2)^2-4sm_W^2$,
$t_{\pm}=\frac{1}{2}[m_t^2+m_W^2-s\pm\lambda^{1/2}]$, and
$\overline{|M_{ren}^A|^2}$ is the renormalized amplitude squared
given by
\begin{eqnarray}
\overline{|M_{ren}^A|^2}&=&\overline{\sum}|M_0^A|^2+
2\mathrm{Re}\overline{\sum}M_0^A[M_{vir}^A+ M_{count}^A]^{\dag},
\end{eqnarray}
where the colors and spins of the outgoing particles have been
summed over, and the colors and spins of incoming ones have been
averaged over.

The total cross section at the LHC is obtained by convoluting the
partonic cross section with the parton distribution functions (PDFs)
$G_{g,b/p}$ in the proton:
\begin{eqnarray}\label{crosssection}
\sigma&=&\int_{\tau_0}^1\mathrm{d}x_1\int_{\tau_0/x_1}^1
\mathrm{d}x_2[G_{g/p}(x_1,\mu_f)
G_{b/p}(x_2,\mu_f)+(x_1\leftrightarrow x_2)]\hat{\sigma}(\tau S),
\end{eqnarray}
where $\mu_f$ is the factorization scale and $S=(P_1+P_2)^2$, $P_1$
and $P_2$ are the four-momentum of the incident hadrons,
$\tau_0=\frac{(m_W+m_t)^2}{S}$, $\tau=x_1x_2$, and $x_1$, $x_2$ are
the longitudinal momentum fractions of initial partons in the
hadrons.

\subsection{s-channel and t-channel}
For convenience, we first define the reduced standard matrix
elements $F_i$ as follows:
\begin{eqnarray}
F_{1,2}&=&\bar{v}(k_2)P_{R,L}\gamma^{\mu}u(k_1)
\bar{u}(k_3)P_R\gamma_{\mu}v(k_4),\notag\\
F_{3,4}&=&\bar{v}(k_2)P_{R,L}u(k_1)
\bar{u}(k_3)P_R\not{k_1}v(k_4),\notag\\
F_{5,6}&=&\bar{u}(k_3)P_{R,L}v(k_4)
\bar{v}(k_2)P_R\not{k_3}u(k_1),\notag\\
F_{7,8}&=&\bar{v}(k_2)P_{R,L}u(k_1)
\bar{u}(k_3)P_Lv(k_4),\notag\\
F_9&=&\bar{v}(k_2)P_R\gamma^{\mu}u(k_1)
\bar{u}(k_3)P_L\gamma_{\mu}v(k_4),
\end{eqnarray}
which will appear in the amplitudes of s-channel and t-channel
below.

For s-channel, the diagrams which contribute to the LO amplitude
$M_0^s$ are shown in Fig.\ref{tree} (c). The LO amplitude $M_0^s$ is
\begin{eqnarray}\label{stree}
M_0^s&=&-\sum_{q=u,c\atop q'=d,s,b}\frac{2\pi\alpha
V_{tb}V_{qq'}^*}{\sin^2\theta_W(s-m_W^2)} F_1,
\end{eqnarray}

The virtual corrections $M_{vir}^s$ contains the radiative
corrections from the one-loop vertex diagrams, which are shown in
Fig.\ref{stloop} (a) and (b), and we can write $M_{vir}^s$ as:
\begin{eqnarray}\label{svir}
M_{vir}^s&=&\sum_{m=1}^9f_m^sF_m,
\end{eqnarray}
where $f_m^s(m=1,2,\cdots,9)$ are form factors, which are given
explicitly in Appendix.

The corresponding counterterm can be written as
\begin{eqnarray}\label{scount}
M_{count}^s&=&-\sum_{q=u,c\atop q'=d,s,b}\frac{\pi\alpha
V_{tb}V_{qq'}^*}{\sin^2\theta_W(s-m_W^2)}(\delta Z_L^{bb}+\delta
Z_L^{tt}+\delta Z_L^{qq}+\delta Z_L^{q'q'})F_1,
\end{eqnarray}
where the expressions of $\delta Z_L^{ii}(i=u,d,s,c,b,t)$ are shown
in Eq.(\ref{RC}).

According to the crossing symmetry, we have the similar expressions
in t-channel as in s-channel. We replace the variable $s$ by $t$ in
Eq.(\ref{stree}), (\ref{svir}) and (\ref{scount}), then use the
different summation over quark flavors, and change the indices of
the quarks in the initial and final states. For example,
\begin{eqnarray}
M^t_0&=&M^s_0(s\rightarrow t,\sum_{q,q'}\rightarrow \sum_{\{qq'\}})
=-\sum_{\{qq'\}}\frac{4\pi\alpha
V_{tb}V_{qq'}^*}{\sin^2\theta_W(t-m_W^2)}F_1',\notag\\
M_{vir}^t&=&M_{vir}^s(s\rightarrow
t,\sum_{q,q'}\rightarrow\sum_{\{qq'\}})
=\sum_{m=1}^9f_m^t(s\rightarrow
t,\sum_{q,q'}\rightarrow\sum_{\{qq'\}})
F_m'\notag\\
M^t_{count}&=&M^s_{count}(s\rightarrow t,\sum_{q,q'}\rightarrow
\sum_{\{qq'\}})\notag\\
&=&-\sum_{\{qq'\}}\frac{2\pi\alpha
V_{tb}V_{qq'}^*}{\sin^2\theta_W(t-m_W^2)}(\delta Z_L^{qq}+\delta
Z_L^{q'q'}+\delta Z_L^{bb}+\delta Z_L^{tt})F_1',
\end{eqnarray}
with
\begin{eqnarray*}
F_m'&\equiv&F_m(v(k_4)\rightarrow u(k_2),\bar{v}(k_2)\rightarrow
u(k_3)).
\end{eqnarray*}
Here the index pair $\{qq'\}$ takes on the flavors $\{ud, us, ub,
cd, cs, cb\}$. The other formulas for cross sections are the same as
in associated production.

\section{Numerical Results and discussions}

In this section, we present the numerical results for the SUSY QCD
corrections to the three channels of single top production at the
LHC. For comparison, we also present the numerical results for
s-channel at the Tevatron\cite{li}. In our numerical calculations,
we use the following set of the SM parameters
 \cite{zhao14}:
$$m_t = 175\mathrm{GeV}, \qquad \alpha_{ew}(M_W) = 1/128, \qquad
 \alpha_s(M_Z)=0.118,$$
and all light quark masses are set to be zero, and the CKM matrix
elements are taken to be the values shown in Ref.\cite{zhao14}.

The running QCD coupling $\alpha_s(Q)$ is evaluated at two-loop
order\cite{pdg}, and the CTEQ6M PDFs\cite{zhao16} are used
throughout this paper to calculate cross sections. For simplicity,
we neglect the b-quark mass. We choose $\mu_r=\mu_f=m_t+m_W$ for the
renormalization and factorization scales in associated production
and choose $\mu_r=\mu_f=m_t$ for the renormalization and
factorization scales in the other two channels.

Besides, the values of the MSSM parameters taken in our numerical
calculations are constrained within the minimal supergravity
scenario(mSUGRA)  \cite{zhao19}, in which there are only five free
input parameters at the grand unification where $M_{1/2}, M_0, A_0,
\tan\beta$ and the sign of $\mu$, where $M_{1/2}, M_0, A_0$ are,
respectively, the universal gaugino mass, scalar mass, and the
trilinear soft breaking parameter in the superpotential. Given these
parameters, all the MSSM parameters at the weak scale are determined
in the mSUGRA scenario by using the program package SUSPECT 2.3
 \cite{zhao20}, where we set $A_0=-200\mathrm{GeV}$ and
$\mu>0$.

\subsection{Associated Production}

We define the K factor as the ratio of the SUSY QCD corrected cross
sections to LO total cross sections, calculated using the CTEQ6M
PDFs. Fig.\ref{gb2twm0150} shows the K factors as functions of
$M_{\tilde{g}}$ ($M_{1/2}$) for the associated production process
$pp\rightarrow tW$ at the LHC for $\tan\beta=5,20$ and $35$,
respectively. From Fig.\ref{gb2twm0150}, we can see that the
differences among the results are small for different $\tan\beta$,
and K factors increase with the increasing $M_{\tilde{g}}$ for small
$M_{\tilde{g}}(\lesssim 160\mathrm{GeV})$, while decrease with the
increasing $M_{\tilde{g}}$ for large $M_{\tilde{g}}(\gtrsim
160\mathrm{GeV})$, and, in general, the K factors are about $1.06$.

In Fig.\ref{gb2twmhf40} we show the dependence of the K factors on
$M_{\tilde{t}_1}$($M_0$) for different $\tan\beta$.
Fig.\ref{gb2twmhf40} shows that the K factors have the similar
behaviors as those shown in Fig.\ref{gb2twm0150}, and are also about
$1.06$ in general.

To compare with Fig.\ref{gb2twmhf40}, we present
Fig.\ref{gb2twmhf70} which takes the similar parameters as those
used in Fig.\ref{gb2twmhf40} but $M_{1/2}=70\mathrm{GeV}$, where the
gluino mass $M_{\tilde{g}}$ lies in the range
$220\mathrm{GeV}\lesssim M_{\tilde{g}}\lesssim250\mathrm{GeV}$ for
all values of $M_0$ and $\tan\beta$ we assumed here.
Fig.\ref{gb2twmhf70} shows that SUSY QCD corrections are not
sensitive to $\tan\beta$, which is consistent with
Fig.\ref{gb2twmhf40}.

Fig.\ref{gb2twtb5m0} gives the K factors as functions of
$M_{\tilde{g}}$($M_{1/2}$) for different $M_0$, assuming
$\tan\beta=5$. In Fig.\ref{gb2twtb5m0} we can see that there are
large differences between different $M_0$ when $M_{\tilde{g}}$
becomes small, but these curves approach each other when
$M_{\tilde{g}}$ becomes large because of the decoupling of heavy
gluino($M_{\tilde{g}}\gtrsim450\mathrm{GeV}$). The K factors are
about $1.06$ for $M_{\tilde{g}}\lesssim500\mathrm{GeV}$, and become
small slowly with increasing $M_{\tilde{g}}$, but the K factors
decrease rapidly when $M_{\tilde{g}}\lesssim150\mathrm{GeV}$ for
$M_0=150\mathrm{GeV}$.

In Fig.\ref{gb2twtb5mhf} we present the K factors as functions of
$M_{\tilde{t}_1}$($M_0$), assuming $\tan\beta=5$, and $M_{1/2}=40,
70$ and $100\mathrm{GeV}$, respectively. From Fig.\ref{gb2twtb5mhf},
we find the similar results as those shown in Fig.\ref{gb2twtb5m0},
i.e. the K factors are about $1.06$ for most values of
$M_{\tilde{t}_1}$ considered, and become small slowly with
increasing $M_{\tilde{t}_1}$.

In Fig.\ref{scalegb2tw} we present the LO and the SUSY QCD corrected
cross sections as functions of renormalization and factorization
scales $\mu/\mu_0$($\mu_f=\mu_r=\mu, \mu_0=m_t+m_W$), respectively,
assuming $\tan\beta=5$, $M_0=150\mathrm{GeV}$,
$M_{1/2}=70\mathrm{GeV}$, $A_0=-200\mathrm{GeV}$ and $\mu>0$. This
figure shows that the scale dependence of the SUSY QCD corrected
total cross section is a little larger than that of the LO cross
section because of the running effects of the extra $\alpha_s$ in
SUSY QCD corrections. We can recover the LO results of scale
dependence by dividing $\alpha_s$ in the SUSY QCD corrections. After
comparison with the NLO QCD corrections\cite{myself2}, we can see
that if the NLO QCD corrections are also included,
$\mathcal{O}(\alpha_s)$ corrections still improve the scale
dependence.

\subsection{s-channel and t-channel}

For the s-channel process $pp\rightarrow t\bar{b}$, in
Fig.\ref{qq2tbmLHC} (a)-(b), we display the K factors as functions
of $M_{\tilde{g}}$($M_{1/2}$) and $M_{\tilde{t}_1}$($M_0$),
respectively, assuming $M_0=150,300\mathrm{GeV}$ and
$M_{1/2}=40,70,100\mathrm{GeV}$. Fig.\ref{qq2tbmLHC} shows that the
K factors are about $1.01$ for some favorable parameters, otherwise,
the K factors approach the unit value.

In Fig.\ref{qq2tbtbLHC} (a)-(b), the K factors are plotted as
functions of $M_{\tilde{g}}$($M_{1/2}$) and
$M_{\tilde{t}_1}$($M_0$), respectively, assuming $\tan\beta=5, 20,
35$. Fig.\ref{qq2tbtbLHC} shows that the K factors can reach $1.01$
for light $M_{\tilde{g}}$ and $M_{\tilde{t}_1}$, respectively, and
are not sensitive to $\tan\beta$.

In Fig.\ref{qq2tbmtev}, we display the K factors as functions of
$M_{\tilde{t}_1}$ for the s-channel process $p\bar{p}\rightarrow
t\bar{b}$ at the Tevatron Run II. We find that the K factors can
reach $1.01$ for small $M_{\tilde{t}_1}$, and decrease rapidly as
$M_{\tilde{t}_1}$ increases. These results are consistent with that
shown in Fig.3 of Ref.\cite{li}, where the relevant parameters
assumed are the same as ones used in our numerical calculations.

For the t-channel process $pp\rightarrow qt$ at the LHC,
Fig.\ref{qb2qtm} shows the K factors as functions of $M_{\tilde{g}}$
for $\tan\beta=5$ and $M_0=150, 300\mathrm{GeV}$, respectively. From
Fig.\ref{qb2qtm}, we find that the SUSY QCD corrections decrease the
total cross sections and the K factors approach the unit value for
all parameters assumed here, which means that SUSY QCD corrections
are negligible.

In Fig.\ref{scaleqq2tb} and Fig.\ref{scaleqb2qt} we present the LO
and the SUSY QCD corrected cross sections as functions of
renormalization and factorization scales
$\mu/\mu_0$($\mu_f=\mu_r=\mu, \mu_0=m_t$) for both s-channel and
t-channel, respectively, assuming $\tan\beta=5$,
$M_0=150\mathrm{GeV}$, $M_{1/2}=70\mathrm{GeV}$,
$A_0=-200\mathrm{GeV}$ and $\mu>0$. Since the SUSY QCD corrections
to the two channels are very small, they do not affect the scale
dependence of the LO results obviously.

\section{Conclusion}
In conclusion, we have calculated the SUSY QCD corrections to the
total cross sections for single top production at the Tevatron and
the LHC in the MSSM. Our results show that for the s-channel and
t-channel, the SUSY QCD corrections are at most about $1\%$, but for
the associated production process $pp\rightarrow tW$, the SUSY QCD
corrections increase the total cross sections significantly, which
can reach about $6\%$ for most values of the parameters, and the
SUSY QCD corrections should be taken into consideration in the
future high precision experimental analysis for top physics at the
LHC.

\begin{acknowledgments}
This work was supported in part by the
National Natural Science Foundation of China, under grants
No.~10421503 and No.~10575001, and the Key Grant Project of Chinese
Ministry of Education under grant No.~305001.
\end{acknowledgments}

\section*{Appendix}

In this Appendix, we will list the explicit expressions of the
non-zero form factors of associated production and s-channel. For
simplicity, we first define the abbreviations for Passarino-Veltman
functions  \cite{myself5} below.
$$B_0^s=B_0(s, M_{\tilde{g}}^2, M_{\tilde{t}_i}^2),\qquad
B_1^s=B_1(s, M_{\tilde{g}}^2, M_{\tilde{t}_i}^2),\qquad
{B'}_0^s=B_0(s, M_{\tilde{g}}^2, M_{\tilde{b}_i}^2),$$
$$B_0^t=B_0(t, M_{\tilde{g}}^2, M_{\tilde{t}_i}^2),\qquad
B_1^t=B_1(t, M_{\tilde{g}}^2, M_{\tilde{t}_i}^2),$$
$$C_{ijk\cdots}^a=C_{ijk\cdots}(m_t^2, s, m_W^2, M^2_{\tilde{t}_i},
M_{\tilde{g}}^2,M^2_{\tilde{b}_j}),$$
$$C_{ijk\cdots}^b=C_{ijk\cdots}(0, t, m_W^2, M_{\tilde{b}_i}^2,
M_{\tilde{g}}^2,M_{\tilde{t}_j}^2),$$
$$C_{ijk\cdots}^c=C_{ijk\cdots}(m_t^2, t, 0, M_{\tilde{g}}^2,
M_{\tilde{t}_i}^2,M_{\tilde{g}}^2),$$
$$C_{ijk\cdots}^d=C_{ijk\cdots}(m_t^2, t, 0, M_{\tilde{t}_i}^2,
M_{\tilde{g}}^2,M_{\tilde{t}_i}^2),$$
$$C_{ijk\cdots}^e=C_{ijk\cdots}(m_t^2, t, 0, M_{\tilde{g}}^2,
M_{\tilde{t}_i}^2,M_{\tilde{t}_i}^2),$$
$$C_{ijk\cdots}^f=C_{ijk\cdots}(0, s, 0, M_{\tilde{g}}^2,
M_{\tilde{b}_i}^2,M_{\tilde{g}}^2),$$
$$C_{ijk\cdots}^g=C_{ijk\cdots}(0, s, 0, M_{\tilde{b}_i}^2,
M_{\tilde{g}}^2,M_{\tilde{b}_i}^2),$$
$$C_{ijk\cdots}^h=C_{ijk\cdots}(0, u, m_t^2, M^2_{\tilde{g}},
M^2_{\tilde{b}_i},M^2_{\tilde{t}_j}),$$
$$C_{ijk\cdots}^i=C_{ijk\cdots}(m_W^2, m_t^2, s, M_{\tilde{b}_i}^2,
M_{\tilde{t}_j}^2,M_{\tilde{g}}^2),$$
$$C^A_{ijk\cdots}=C_{ijk\cdots}(0, s, 0, M_{\tilde{g}}^2,
M_{\tilde{u}_i}^2, M_{\tilde{d}_j}^2),$$
$$C^B_{ijk\cdots}=C_{ijk\cdots}(m_t^2, s, 0, M_{\tilde{g}}^2,
M_{\tilde{t}_i}^2, M_{\tilde{b}_j}^2),$$
$$D_{ijk\cdots}^a=D_{ijk\cdots}(0, m_W^2, m_t^2, 0, t, s,
M_{\tilde{g}}^2, M_{\tilde{b}_i}^2, M_{\tilde{t}_j}^2,
M_{\tilde{g}}^2),$$
$$D_{ijk\cdots}^b=D_{ijk\cdots}(0, s, m_W^2, u,
0, m_t^2, M_{\tilde{b}_i}^2, M_{\tilde{g}}^2, M_{\tilde{b}_i}^2,
M_{\tilde{t}_j}^2),$$
$$D_{ijk\cdots}^c=D_{ijk\cdots}(m_t^2, t,
m_W^2, u, 0, 0, M_{\tilde{t}_i}^2, M_{\tilde{g}}^2,
M_{\tilde{t}_i}^2, M_{\tilde{b}_j}^2).$$

For s-channel, we have
$$
f_1^s=-\sum_{q=u,c\atop
q'=d,s,b}\sum_{i,j=1}^2\frac{8\alpha\alpha_sV_{tb}V_{qq'}^*}{3\sin^2\theta_W(s-m_W^2)}
(U_{i1}^qU_{j1}^{q'}U_{i1}^{q*}U_{j1}^{q'*}
C^A_{00}+U_{i1}^tU_{j1}^bU_{i1}^{t*} U_{j1}^{b*}C^B_{00}),$$$$
f_2^s=-\sum_{q=u,c\atop
q'=d,s,b}\sum_{i,j=1}^2\frac{8\alpha\alpha_sV_{tb}V_{qq'}^*}{3\sin^2\theta_W(s-m_W^2)}
U_{i2}^qU_{j1}^{q'}U_{i1}^{q*}U_{j2}^{q'*}C^A_{00},$$$$
f_3^s=-\sum_{q=u,c\atop
q'=d,s,b}\sum_{i,j=1}^2\frac{8\alpha\alpha_sV_{tb}V_{qq'}^*}{3\sin^2\theta_W(s-m_W^2)}
M_{\tilde{g}}U_{i2}^qU_{j1}^{q'}U_{i1}^{q*}U_{j1}^{q'*}(C^A_0
+C^A_1+C^A_2),$$$$ f_4^s=-\sum_{q=u,c\atop
q'=d,s,b}\sum_{i,j=1}^2\frac{8\alpha\alpha_sV_{tb}V_{qq'}^*}{3\sin^2\theta_W(s-m_W^2)}
M_{\tilde{g}}U_{i1}^qU_{j1}^{q'}U_{i1}^{q*}U_{j2}^{q'*}(C^A_0+
C^A_1+C^A_2),$$$$ f_5^s=-\sum_{q=u,c\atop
q'=d,s,b}\sum_{i,j=1}^2\frac{8\alpha\alpha_sV_{tb}V_{qq'}^*}{3\sin^2\theta_W(s-m_W^2)}
[M_{\tilde{g}}U_{i1}^tU_{j2}^b
U_{i1}^{t*}U_{j1}^{b*}(C^B_0+C^B_1+C^B_2)$$$$ +m_tU_{i1}^tU_{j2}^b
U_{i2}^{t*}U_{j1}^{b*}(C^B_1+C^B_{11}+C^B_{12})],$$$$
f_6^s=-\sum_{q=u,c\atop
q'=d,s,b}\sum_{i,j=1}^2\frac{8\alpha\alpha_sV_{tb}V_{qq'}^*}{3\sin^2\theta_W(s-m_W^2)}
[m_tU_{i1}^tU_{j1}^bU_{i1}^{t*}U_{j1}^{b*}(C^B_1+C^B_{11}+
C^B_{12})$$$$ +M_{\tilde{g}}U_{i1}^tU_{j1}^bU_{i2}^{t*}
U_{j1}^{b*}(C^B_0+C^B_1+C^B_2)],$$$$ f_7^s=\sum_{q=u,c\atop
q'=d,s,b}\sum_{i,j=1}^2\frac{4\alpha\alpha_sV_{tb}V_{qq'}^*}{3\sin^2\theta_W(s-m_W^2)}
m_tM_{\tilde{g}}U_{i2}^qU_{j1}^{q'}U_{i1}^{q*}U_{j1}^{q'*}(C^A_0+
2C^A_2),$$$$ f_8^s=\sum_{q=u,c\atop
q'=d,s,b}\sum_{i,j=1}^2\frac{4\alpha\alpha_sV_{tb}V_{qq'}^*}{3\sin^2\theta_W(s-m_W^2)}
m_tM_{\tilde{g}}U_{i1}^qU_{j1}^{q'}U_{i1}^{q*}U_{j2}^{q'*}(C^A_0
+2C^A_2),$$$$ f_9^s=-\sum_{q=u,c\atop
q'=d,s,b}\sum_{i,j=1}^2\frac{8\alpha\alpha_sV_{tb}V_{qq'}^*}{3\sin^2\theta_W(s-m_W^2)}
C^B_{00}U_{i1}^tU_{j2}^bU_{i2}^{t*}U_{j1}^{b*}.
$$

\begin{figure}
\begin{center}
\scalebox{0.8}{\includegraphics[110,460][800,710]{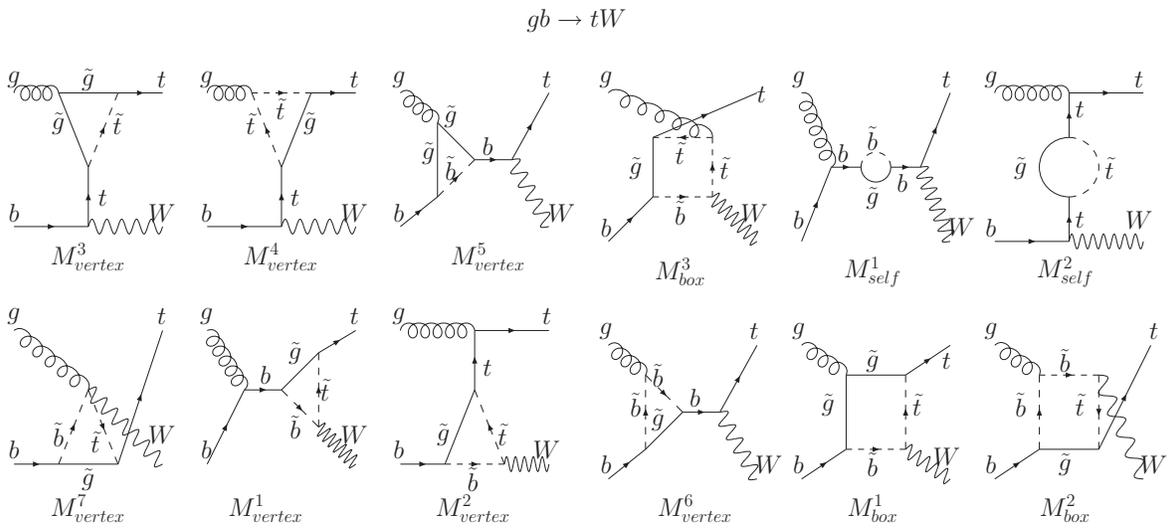}}
\end{center}
\caption{The Feynman diagrams for SUSY QCD corrections of associate
production.}\label{gb2tw}
\end{figure}

\begin{figure}
\begin{center}
\scalebox{0.6}{\includegraphics[120,420][620,720]{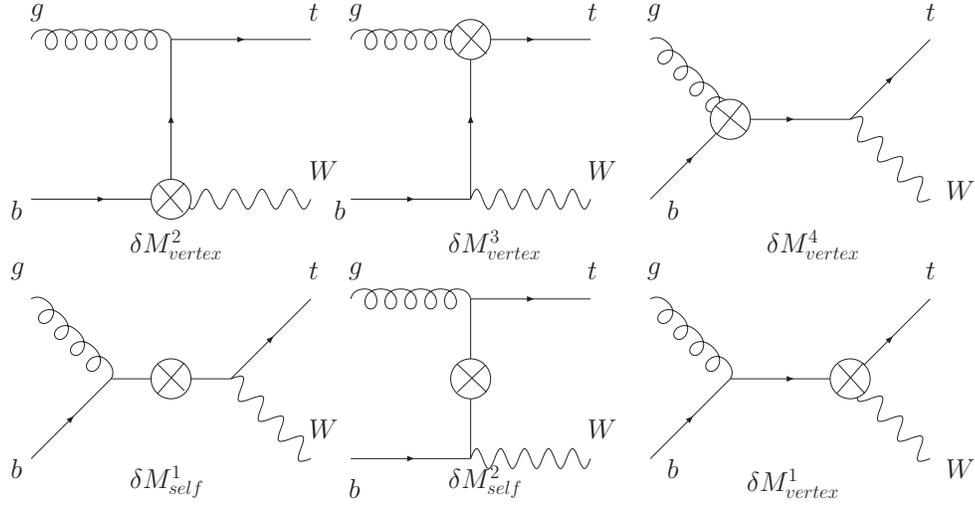}}
\end{center}
\caption{The counter term diagrams for associated production.}
\label{counterterm}
\end{figure}

\begin{figure}
\scalebox{0.6}{\includegraphics[70,280][575,700]{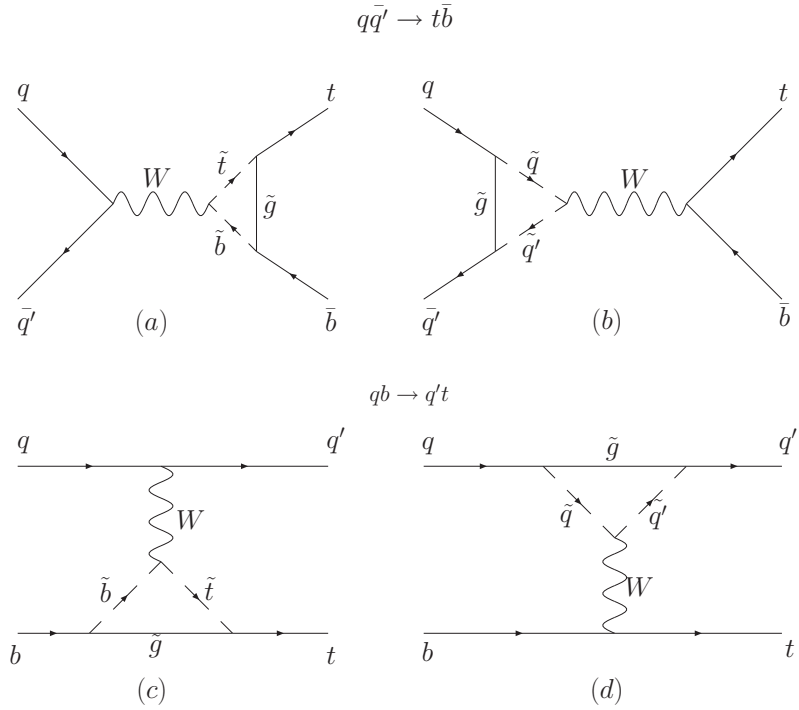}}
\caption{The Feynman diagrams for s-channel and t-channel loop
corrections, respectively.}\label{stloop}
\end{figure}

\begin{figure}
\scalebox{1.1}{\includegraphics[30,25][300,230]{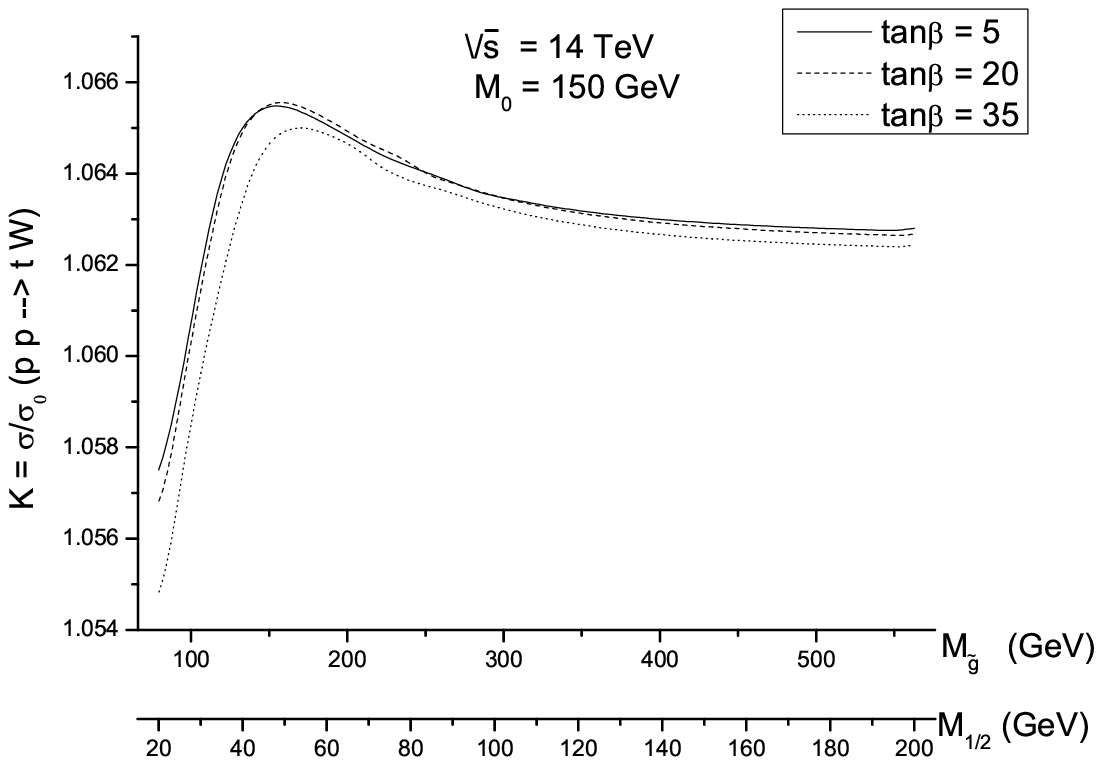}}
\caption{The K factors as functions of $M_{\tilde{g}}$ or $M_{1/2}$
for $pp\rightarrow tW$ at the LHC, where different curves correspond
different $\tan\beta$, assuming: $M_0=150\mathrm{GeV}$,
$A_0=-200\mathrm{GeV}$ and $\mu>0$.} \label{gb2twm0150}
%
\scalebox{1.1}{\includegraphics[30,25][300,260]{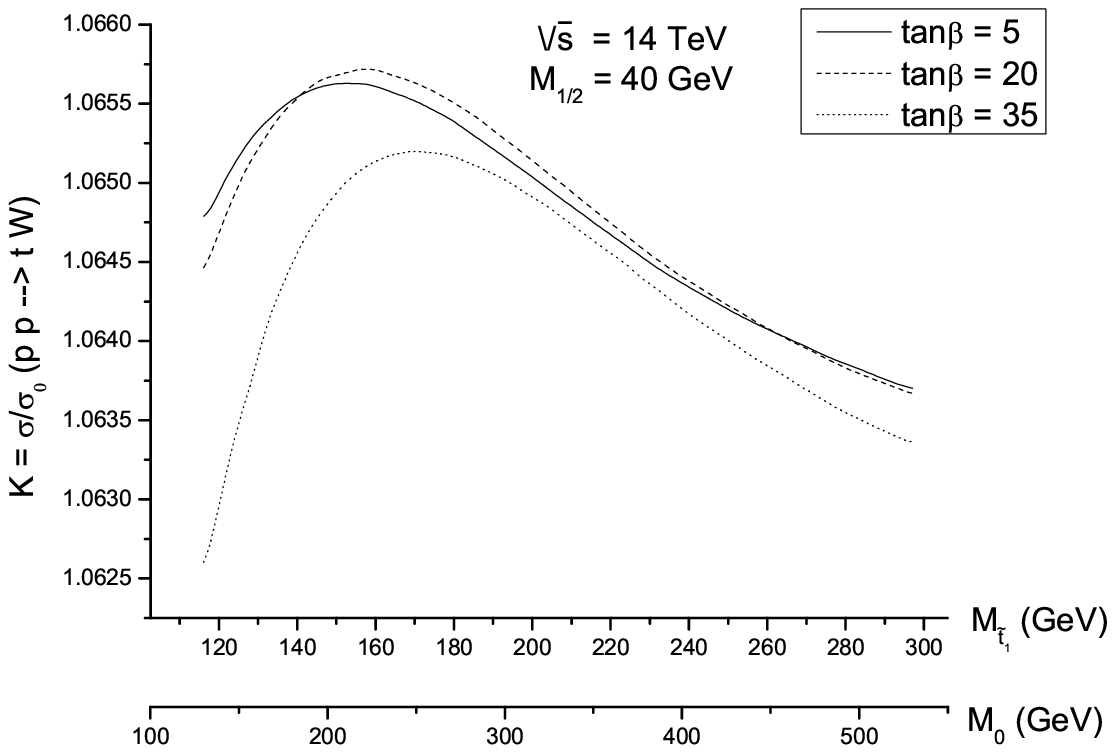}}
\caption{The K factors as functions of $M_{\tilde{t}_1}$ or $M_{0}$
for $pp\rightarrow tW$ at the LHC, where different curves correspond
different $\tan\beta$, assuming: $M_{1/2}=40\mathrm{GeV}$,
$A_0=-200\mathrm{GeV}$ and $\mu>0$.} \label{gb2twmhf40}
\end{figure}

\begin{figure}
\scalebox{1.1}{\includegraphics[30,25][300,230]{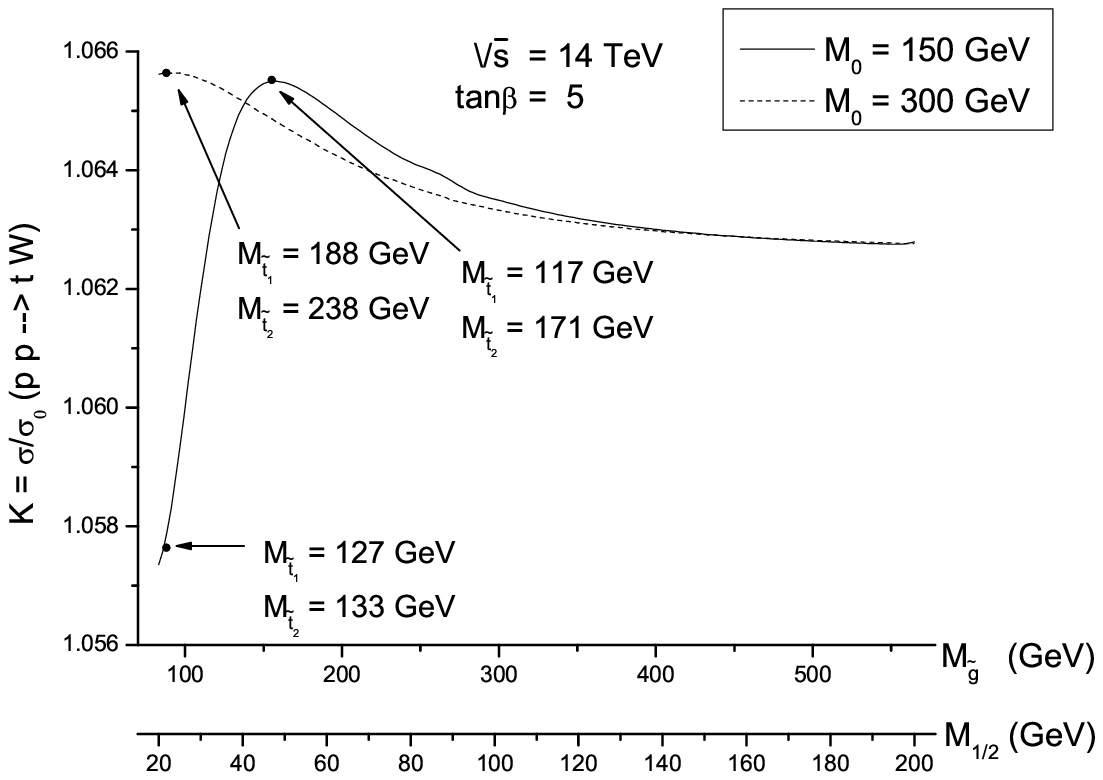}}
\caption{The K factors as functions of $M_{\tilde{g}}$ or $M_{1/2}$
for $pp\rightarrow tW$ at the LHC, where different curves correspond
different $M_0$, assuming: $\tan\beta=5$, $A_0=-200\mathrm{GeV}$ and
$\mu>0$.} \label{gb2twtb5m0}
%
\scalebox{1.1}{\includegraphics[30,25][300,260]{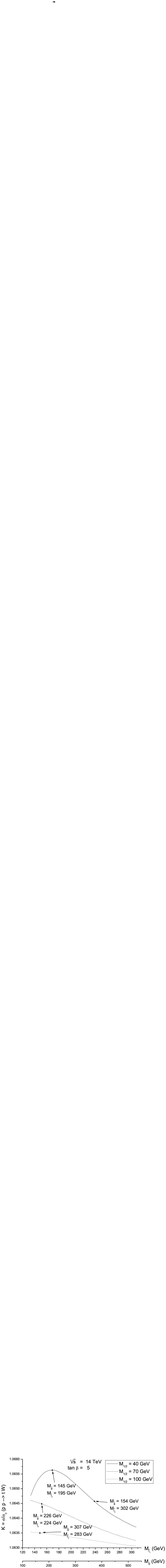}}
\caption{The K factors as functions of $M_{\tilde{t}_1}$ or $M_{0}$
for $pp\rightarrow tW$ at the LHC, where different curves correspond
different $M_{1/2}$, assuming: $\tan\beta=5$, $A_0=-200\mathrm{GeV}$
and $\mu>0$.} \label{gb2twtb5mhf}
\end{figure}

\begin{figure}
\scalebox{1.1}{\includegraphics[30,25][300,230]{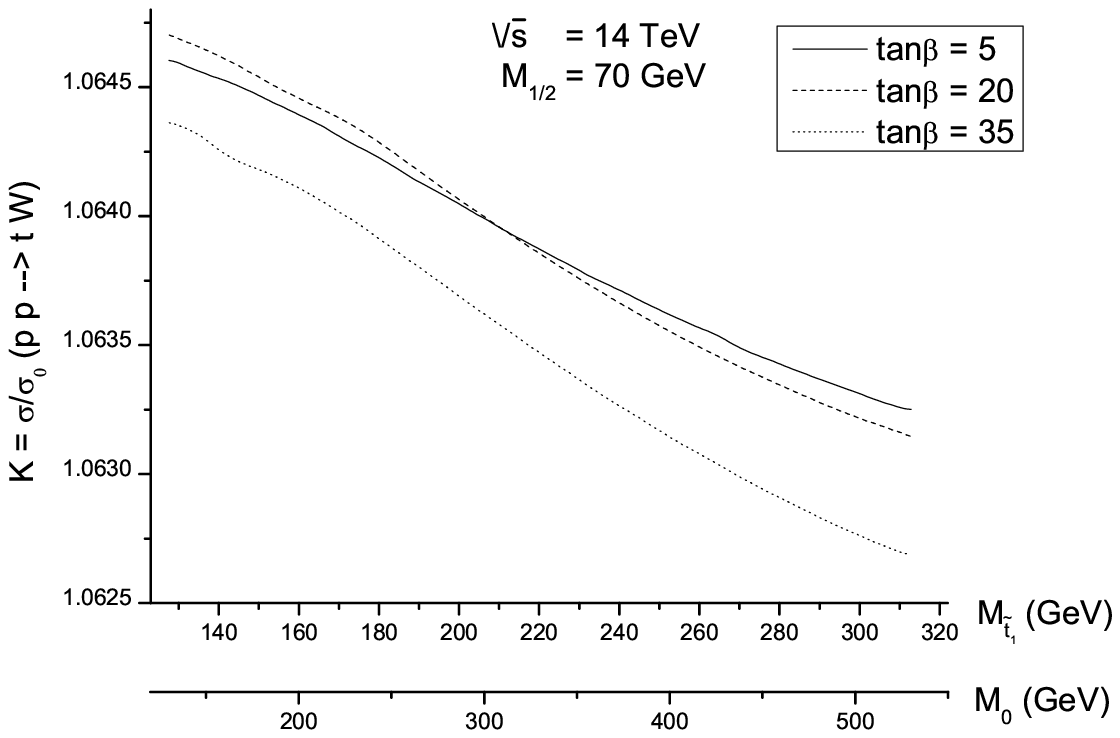}}
\caption{The K factors as functions of $M_{\tilde{t}_1}$ or $M_{0}$
for $pp\rightarrow tW$ at the LHC, where different curves correspond
different $\tan\beta$, assuming: $M_{1/2}=70\mathrm{GeV}$,
$A_0=-200\mathrm{GeV}$ and $\mu>0$.} \label{gb2twmhf70}
\end{figure}

\begin{figure}
\scalebox{1.1}{\includegraphics[30,25][300,260]{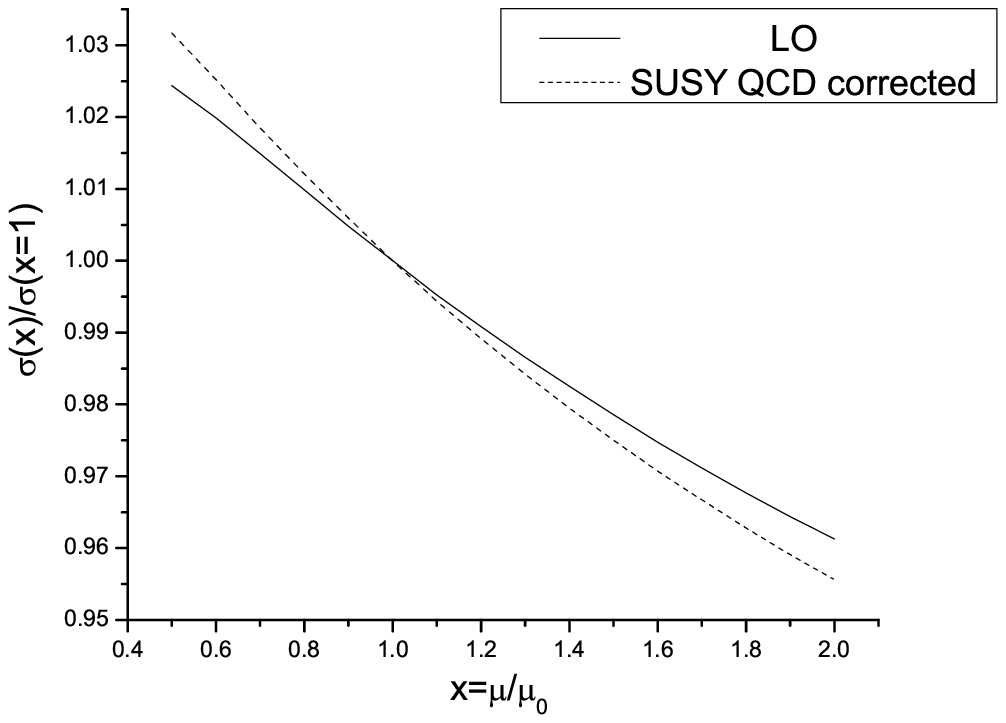}}
\caption{The scale dependence of LO and SUSY QCD corrected cross
sections of associated production at the LHC, we set the
factorization and renormalization scales as $\mu_f=\mu_r=\mu$, and
$\mu_0=m_W+m_t$, assuming: $\tan\beta=5$, $M_0=150\mathrm{GeV}$,
$M_{1/2}=70\mathrm{GeV}$, $A_0=-200\mathrm{GeV}$ and
$\mu>0$.}\label{scalegb2tw}
\end{figure}

\begin{figure}
\scalebox{1.1}{\includegraphics[30,25][300,250]{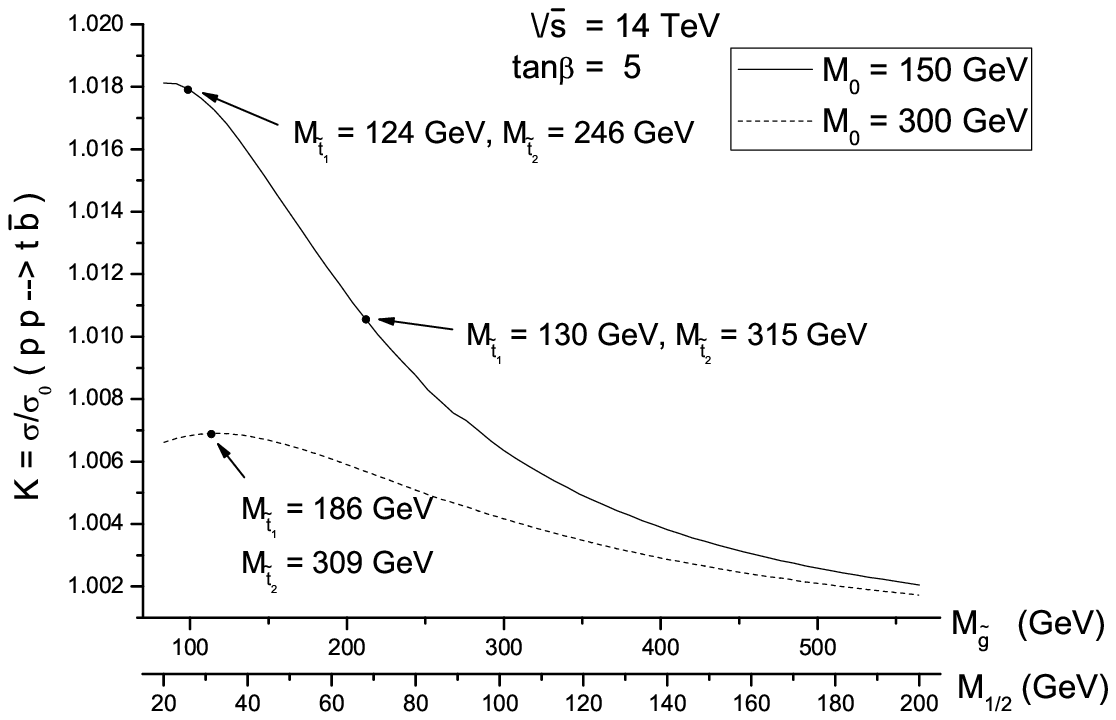}}
\begin{center}
(a)
\end{center}
\scalebox{1.1}{\includegraphics[30,25][300,280]{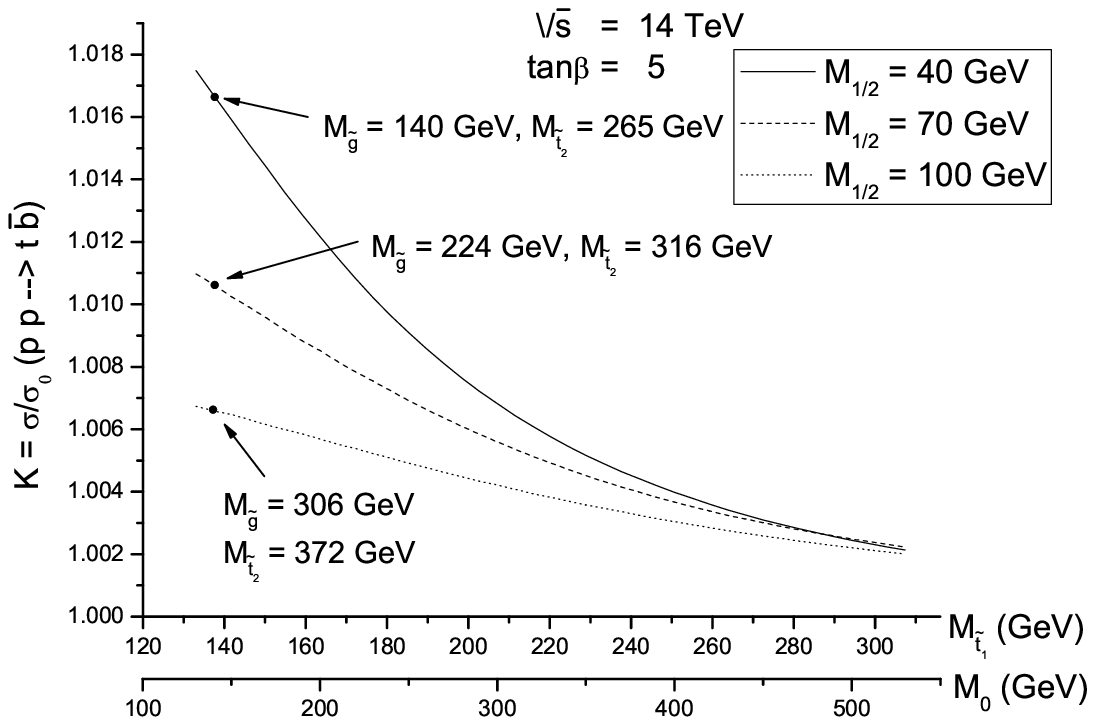}}
\begin{center}
(b)
\end{center}
\caption{The K factors for the $pp\rightarrow t\bar{\mathrm{b}}$ at
the LHC. The variables are (a): $M_{\tilde{g}}$( $M_{1/2}$) and (b):
$M_{\tilde{t}_1}$($M_0$), respectively, assuming: $\tan\beta=5$,
$A_0=-200\mathrm{GeV}$ and $\mu>0$.} \label{qq2tbmLHC}
\end{figure}

\begin{figure}
\scalebox{1.1}{\includegraphics[30,25][300,250]{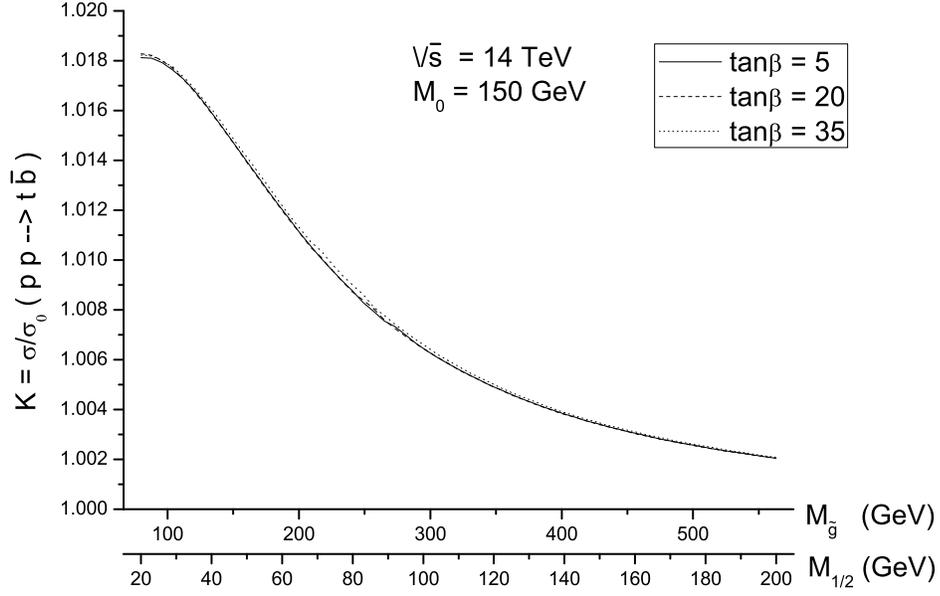}}
\begin{center}
(a)
\end{center}
\scalebox{1.1}{\includegraphics[30,25][300,280]{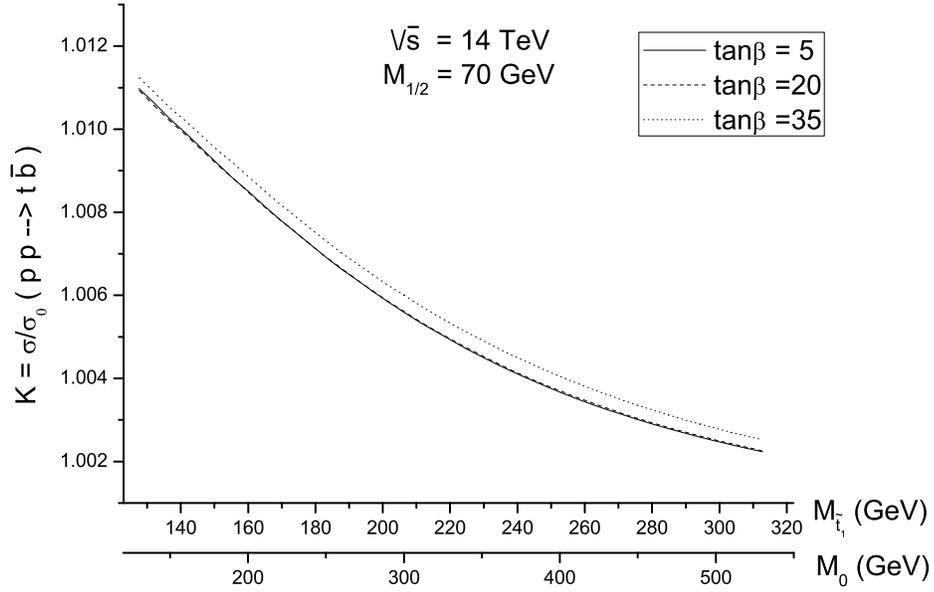}}
\begin{center}
(b)
\end{center}
\caption{The K factors for the $pp\rightarrow t\bar{\mathrm{b}}$ at
the LHC. The variables are (a): $M_{\tilde{g}}$($M_{1/2}$) and (b):
$M_{\tilde{t}_1}$($M_0$), respectively, assuming:
$A_0=-200\mathrm{GeV}$ and $\mu>0$.} \label{qq2tbtbLHC}
\end{figure}

\begin{figure}
\scalebox{1.1}{\includegraphics[30,25][300,240]{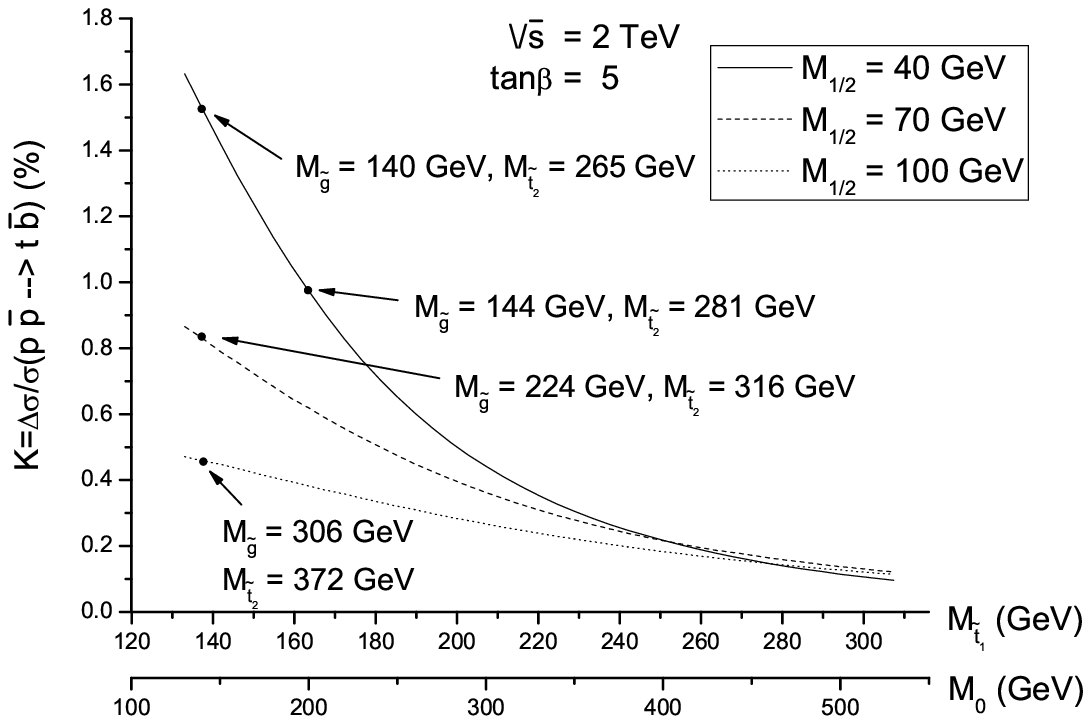}}
\caption{The K factors as functions of $M_{\tilde{t}_1}$($M_0$) for
the $pp\rightarrow t\bar{b}$ at the Tevatron. The graph shows
different $M_{1/2}$, assuming: $\tan\beta=5$, $A_0=-200\mathrm{GeV}$
and $\mu>0$.} \label{qq2tbmtev}
%
\scalebox{1.1}{\includegraphics[30,25][300,270]{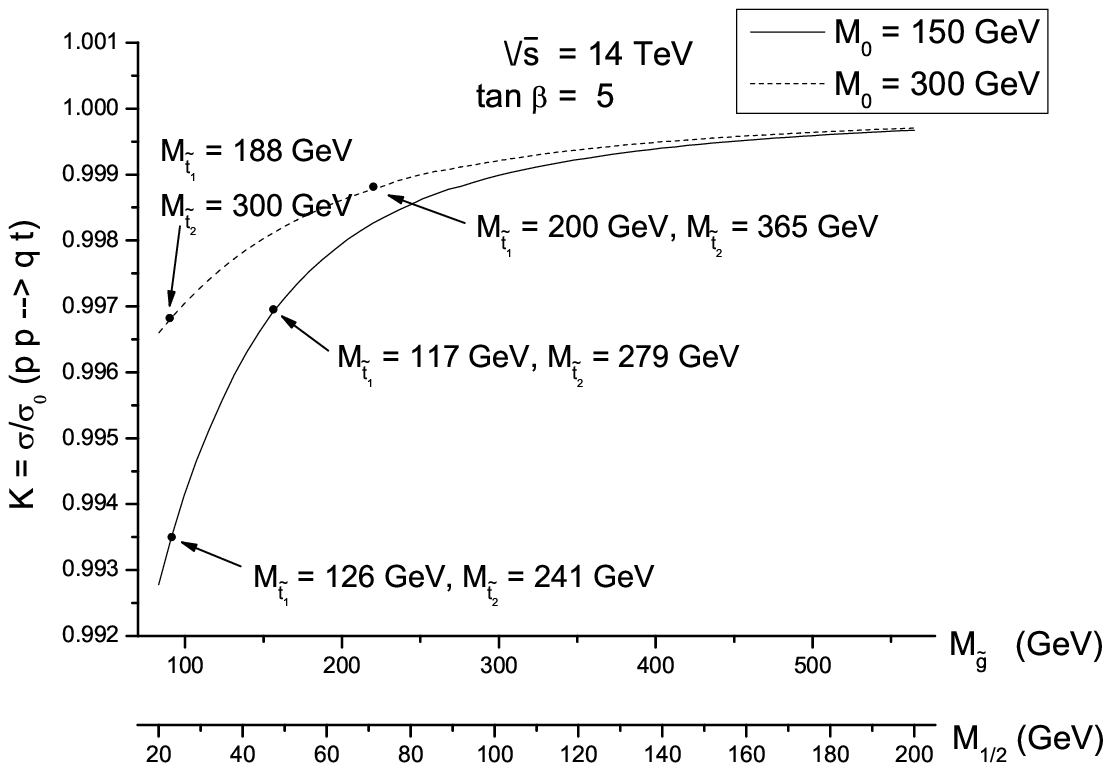}}
\caption{The K factors as functions of $M_{\tilde{g}}$($M_{1/2}$)
for the $pp\rightarrow qt$ at the LHC, the graph shows different
$M_{0}$, assuming: $\tan\beta=5$, $A_0=-200\mathrm{GeV}$ and
$\mu>0$.} \label{qb2qtm}
\end{figure}

\begin{figure}
\scalebox{1.1}{\includegraphics[30,25][300,240]{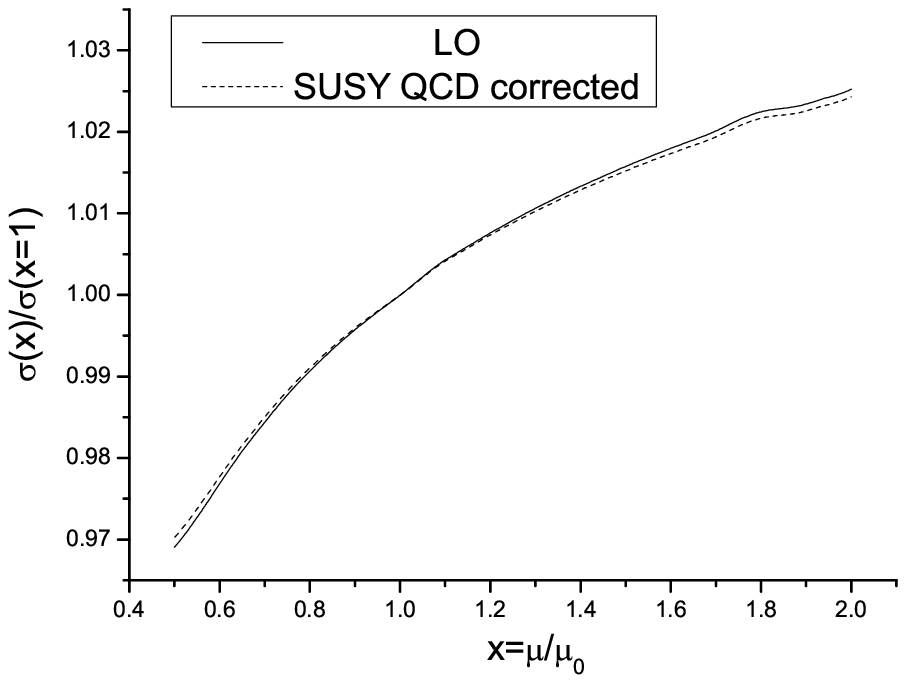}}
\caption{The scale dependence of LO and SUSY QCD corrected cross
sections of s-channel at the LHC, we set the factorization and
renormalization scales as $\mu_f=\mu_r=\mu$, and $\mu_0=m_t$,
assuming: $\tan\beta=5$, $M_0=150\mathrm{GeV}$,
$M_{1/2}=70\mathrm{GeV}$, $A_0=-200\mathrm{GeV}$ and $\mu>0$.}
\label{scaleqq2tb}
%
\scalebox{1.1}{\includegraphics[30,25][300,270]{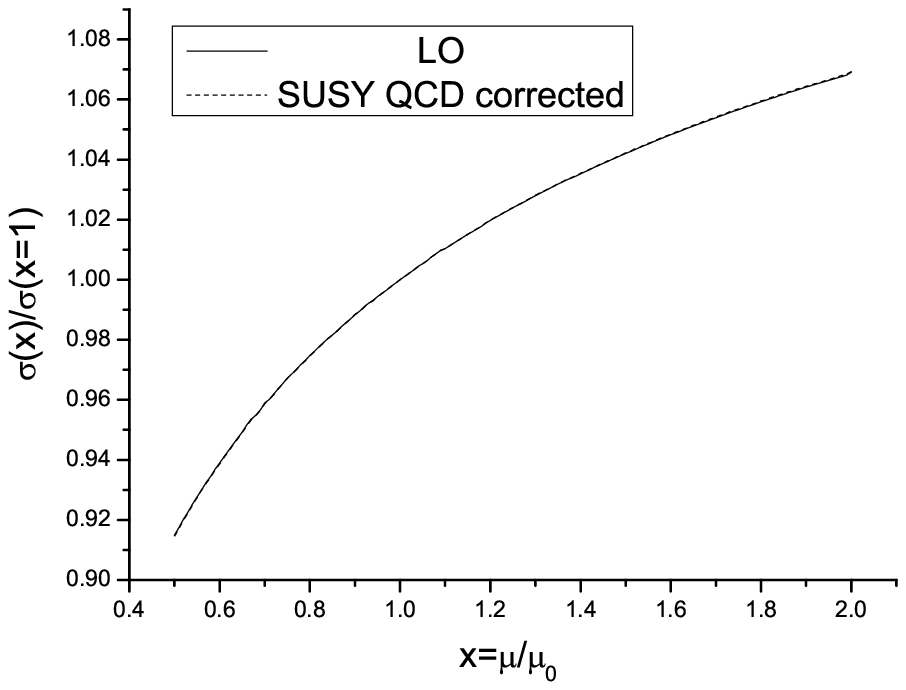}}
\caption{The scale dependence of LO and SUSY QCD corrected cross
sections of t-channel at the LHC, we set the factorization and
renormalization scales as $\mu_f=\mu_r=\mu$, and $\mu_0=m_t$,
assuming: $\tan\beta=5$, $M_0=150\mathrm{GeV}$,
$M_{1/2}=70\mathrm{GeV}$, $A_0=-200\mathrm{GeV}$ and $\mu>0$.}
\label{scaleqb2qt}
\end{figure}

For associated production, we calculate the individual diagram
separately by different types, as shown in Fig.\ref{gb2tw} and
Eq.(\ref{general}). The form factors of the self-energy diagrams are
$$
f_1^{self}=-\frac{2\sqrt{2}\alpha_sg_seV_{tb}}{3\pi
s\sin\theta_W}\sum_i^2 M_{\tilde{g}}U^b_{i2}U^{b*}_{i1}B_0^s,$$$$
f_7^{self}=\frac{\sqrt{2}\alpha_sg_seV_{tb}}{3\pi\sin\theta_W}\sum_i^2
\{\frac{1}{s}U^b_{i1}U^{b*}_{i1}B_1^s-\frac{1}{(t-m_t^2)^2}
[M_{\tilde{g}}m_t(U^t_{i2}U^{t*}_{i1}+
U^t_{i1}U^{t*}_{i2})B_0^t$$$$ +(tU^t_{i1}U^{t*}_{i1}
+m_t^2U^t_{i2}U^{t*}_{i2})B_1^t]\},$$$$
f_8^{self}=\frac{2\sqrt{2}\alpha_sg_seV_{tb}}{3\pi
s\sin\theta_W}\sum_i^2 U^b_{i1}U^{b*}_{i1}B_1^s,$$$$
f_9^{self}=-f_{15}^{self}=-\frac{2\sqrt{2}\alpha_sg_seV_{tb}}{3\pi\sin\theta_W
(t-m_t^2)^2}\sum_i^2[M_{\tilde{g}}m_t(U^t_{i2}U^{t*}_{i1}+
U^t_{i1}U^{t*}_{i2})B_0^t+(tU^t_{i1}U^{t*}_{i1}
+m_t^2U^t_{i2}U^{t*}_{i2})B_1^t],$$$$
f_{10}^{self}=\frac{\sqrt{2}\alpha_sg_seV_{tb}}{3\pi
s\sin\theta_W}\sum_i^2
[M_{\tilde{g}}U^b_{i2}U^{b*}_{i1}B_0^s+2U^b_{i1}U^{b*}_{i1}B_1^s],
$$$$
f_{17}^{self}=\frac{\sqrt{2}\alpha_sg_seV_{tb}}{3\pi\sin\theta_W
(t-m_t^2)^2}\sum_i^2 [M_{\tilde{g}}(m_t^2
U^t_{i1}U^{t*}_{i2}-tU^t_{i1}U^{t*}_{i2})B_0^t
+m_t(m_t^2U^t_{i2}U^{t*}_{i2}-tU^t_{i2}U^{t*}_{i2})B_1^t].
$$

To avoid the very long expressions, we define
\begin{eqnarray*}
f_m^v&\equiv&\sum_{\alpha=1}^7f_m^{v\alpha},
\end{eqnarray*}
as shown in Fig.\ref{gb2tw}, where
$$
f_2^{v1}=f_3^{v1}=-\sum_i^2\sum_j^2
\frac{4\sqrt{2}\alpha_seg_sV_{tb}
U^t_{i1}U^{b*}_{j1}}{3s\pi\sin\theta_W}[m_tU^b_{j2}U^{t*}_{i2}
(C_1^a+C_{11}^a+C_{12}^a)-M_{\tilde{g}}U^b_{j2}U^{t*}_{i1}C_1^a],$$$$
f_7^{v1}=\frac{1}{2}f_8^{v1}=-\frac{1}{2}f_{10}^{v1}=-\sum_i^2\sum_j^2
\frac{4\sqrt{2}\alpha_seg_sV_{tb}
U^t_{i1}U^{b*}_{j1}}{3s\pi\sin\theta_W}U^b_{j1}U^{t*}_{i1}C_{00}^a,$$$$
f_{15}^{v1}=f_{16}^{v1}=-\sum_i^2\sum_j^2
\frac{2\sqrt{2}\alpha_seg_sV_{tb}
U^t_{i1}U^{b*}_{j1}}{3\pi\sin\theta_W}U^b_{j1}U^{t*}_{i1}C_{12}^a,$$$$
f_{18}^{v1}=f_{19}^{v1}=-\sum_i^2\sum_j^2
\frac{2\sqrt{2}\alpha_seg_sV_{tb}
U^t_{i1}U^{b*}_{j1}}{3s\pi\sin\theta_W}[M_{\tilde{g}}U^b_{j1}U^{t*}_{i2}C_1^a
-m_tU^b_{j2}U^{t*}_{i2}(C_1^a+C_{11}^a-C_{12}^a)],$$$$
f_{20}^{v1}=f_{21}^{v1}=-\sum_i^2\sum_j^2\frac{4\sqrt{2}\alpha_seg_sV_{tb}
U^t_{i1}U^{b*}_{j1}}{3s\pi\sin\theta_W}[m_tU^b_{j1}U^{t*}_{i1}(C_1^a+
C_{11}^a+C_{12}^a) -M_{\tilde{g}}U^b_{j2}U^{t*}_{i1}C_1^a],$$$$
f_{25}^{v1}=f_{26}^{v1}=\sum_i^2\sum_j^2
\frac{2\sqrt{2}\alpha_seg_sV_{tb}
U^t_{i1}U^{b*}_{j1}}{3\pi\sin\theta_W}U^b_{j2}U^{t*}_{i2}C_{12}^a,$$$$
f_{27}^{v1}=\frac{1}{2}f_{30}^{v1}=-\frac{1}{2}f_{32}^{v1}=-\sum_i^2\sum_j^2
\frac{2\sqrt{2}\alpha_seg_sV_{tb}
U^t_{i1}U^{b*}_{j1}}{3s\pi\sin\theta_W}U^b_{j2}U^{t*}_{i2}C_{00}^a,$$$$
f_{28}^{v1}=f_{29}^{v1}=-\sum_i^2\sum_j^2
\frac{2\sqrt{2}\alpha_seg_sV_{tb}
U^t_{i1}U^{b*}_{j1}}{3s\pi\sin\theta_W}[M_{\tilde{g}}U^b_{j1}U^{t*}_{i2}C_1^a
-m_tU^b_{j1}U^{t*}_{i1}(C_1^a+ C_{11}^a+C_{12}^a)],$$
$$
f_4^{v2}=-2f_{19}^{v2}=\frac{4\sqrt{2}\alpha_seg_sV_{tb}}{3\pi\sin\theta_W(t-m_t^2)}
\sum_i^2\sum_j^2U^t_{j1}U^{b*}_{i1}(m_tU^b_{i2}U^{t*}_{j2}C_{12}^b
+M_{\tilde{g}}U^b_{i2}U^{t*}_{j1}C_1^b),$$
$$
f_7^{v2}=\frac{1}{2}f_9^{v2}=-\frac{1}{2}f_{15}^{v2}=
-\frac{2\sqrt{2}\alpha_seg_sV_{tb}}{3\pi\sin\theta_W(t-m_t^2)}
\sum_i^2\sum_j^2U^t_{j1}U^{b*}_{i1}U^b_{i1}U^{t*}_{j1}C_{00}^b,$$$$
f_{16}^{v2}=\frac{2\sqrt{2}\alpha_seg_sV_{tb}}{3\pi\sin\theta_W}
\sum_i^2\sum_j^2U^t_{j1}U^{b*}_{i1}U^b_{i1}U^{t*}_{j1}C_{12}^b,$$$$
f_{23}^{v2}=-2f_{29}^{v2}=\frac{4\sqrt{2}\alpha_seg_sV_{tb}}{3\pi\sin\theta_W(t-m_t^2)}
\sum_i^2\sum_j^2U^t_{j1}U^{b*}_{i1}(m_tU^b_{i1}U^{t*}_{j1}C_{12}^b
+M_{\tilde{g}}U^b_{i1}U^{t*}_{j2}C_1^b),$$$$
f_{25}^{v2}=-2f_{27}^{v2}=-f_{31}^{v2}=\frac{4\sqrt{2}\alpha_seg_sV_{tb}}
{3\pi\sin\theta_W(t-m_t^2)}
\sum_i^2\sum_j^2U^t_{j1}U^{b*}_{i1}U^b_{i2}U^{t*}_{j2}C_{00}^b,$$$$
f_{26}^{v2}=\frac{2\sqrt{2}\alpha_seg_sV_{tb}}{3\pi\sin\theta_W}
\sum_i^2\sum_j^2U^t_{j1}U^{b*}_{i1}U^b_{i2}U^{t*}_{j2}C_{12}^b,$$
$$
f_7^{v3}+f_7^{v4}=-\frac{1}{2}(f_{15}^{v3}+f_{15}^{v4})=\frac{\alpha_seg_sV_{tb}}
{12\sqrt{2}\pi\sin\theta_W(t-m_t^2)}\sum_i^2
[(18C_{00}^c+2C_{00}^d-9B_0^t-9m_t^2C_1^e)U^t_{i1}U^{t*}_{i1}$$$$
-m_t(9M_{\tilde{g}}U^t_{i1}U^{t*}_{i2}C_0^c
+2M_{\tilde{g}}U^t_{i2}U^{t*}_{i1}C_0^c
+9m_tU^t_{i2}U^{t*}_{i2}C_1^e)],$$$$
f_9^{v3}+f_9^{v4}=\frac{\alpha_seg_sV_{tb}}
{12\sqrt{2}\pi\sin\theta_W(t-m_t^2)}\sum_i^2
[(18m_t^2C_{11}^c+18m_t^2C_{12}^c+36C_{00}^c-18B_0^t
-18tC_{12}^c+4C_{00}^d$$$$ +2m_t^2C_{12}^d+2m_t^2C_{11}^d
-2tC_{12}^d+m_t^2C_1^d)U^t_{i1}U^{t*}_{i1}+2m_t^2(9C_{11}^c+
C_1^d+C_{11}^d)U^t_{i2}U^{t*}_{i2}$$$$
+2m_tM_{\tilde{g}}U^t_{i2}U^{t*}_{i1}(9C_1^e -C_1^d)
+2M_{\tilde{g}}m_tU^t_{i1}U^{t*}_{i2}(9C_1^e -C_1^d)],$$$$
f_{17}^{v3}+f_{17}^{v4}=\frac{\alpha_seg_sV_{tb}}
{12\sqrt{2}\pi\sin\theta_W(t-m_t^2)}\sum_i^2
[m_tU^t_{i1}U^{t*}_{i1}(9B_0^t -18C_{00}^c -2C_{00}^d +9m_t^2C_1^e
-9tC_1^e)$$$$ +9M_{\tilde{g}}U^t_{i1}U^{t*}_{i2}(m_t^2-t)C_0^c
+m_tU^t_{i2}U^{t*}_{i2}(18C_{00}^c -9B_0^t +2C_{00}^d)],$$
$$
f_{24}^{v3}+f_{24}^{v4}=-2(f_{37}^{v3}+f_{37}^{v4})=\frac{\alpha_seg_sV_{tb}}
{3\sqrt{2}\pi\sin\theta_W(t-m_t^2)}\sum_i^2
[M_{\tilde{g}}U^t_{i1}U^{t*}_{i2}(C_1^d -9C_0^c -9C_1^e)$$$$
+m_tU^t_{i2}U^{t*}_{i2}(9C_{12}^c +C_{12}^d)
-m_tU^t_{i1}U^{t*}_{i1}(9C_{11}^c +9C_{12}^c +C_{12}^d +C_1^d
+C_{11}^d+9C_1^e)],$$
$$
f_1^{v5}+f_1^{v6}=-2(f_6^{v5}+f_6^{v6})=-\frac{3\alpha_seg_sV_{tb}}{2\sqrt{2}\pi
\sin\theta_W}\sum_i^2M_{\tilde{g}}U^b_{i2}U^{b*}_{i1}C_0^f,$$$$
f_7^{v5}+f_7^{v6}=-\frac{1}{2}(f_{10}^{v5}+f_{10}^{v6})=-\frac{\alpha_seg_sV_{tb}}{12\sqrt{2}\pi
s\sin\theta_W}\sum_i^2U^b_{i1}U^{b*}_{i1}(9{B'}_0^s-2C_{00}^g-18C_{00}^f),$$$$
f_8^{v5}+f_8^{v6}=-\frac{\alpha_seg_sV_{tb}}{6\sqrt{2}\pi
s\sin\theta_W}\sum_i^2[U^b_{i1}U^{b*}_{i1}(9{B'}_0^s
-2C_{00}^g-18C_{00}^f)+sU^b_{i1}U^{b*}_{i1}(9C_{12}^f+C_{12}^g)],$$$$
f_{38}^{v5}+f_{10}^{v6}=-\frac{\alpha_seg_sV_{tb}}{6\sqrt{2}\pi
s\sin\theta_W}\sum_i^2M_{\tilde{g}}U^b_{i2}U^{b*}_{i1}(C_1^g-9C_1^f-9C_0^f),
$$
$$
f_1^{v7}=-\frac{\alpha_seg_sV_{tb}}{6\sqrt{2}\pi\sin\theta_W}
\sum_i^2\sum_j^2U^t_{j1}U^{b*}_{i1}U^b_{i2}(M_{\tilde{g}}U^{t*}_{j1}C_0^h+
m_tU^{t*}_{j2}C_2^h),$$$$
f_{22}^{v7}=-\frac{\alpha_seg_sV_{tb}}{6\sqrt{2}\pi\sin\theta_W}
\sum_i^2\sum_j^2U^t_{j1}U^{b*}_{i1}U^b_{i1}(m_tU^{t*}_{j1}C_2^h+
M_{\tilde{g}}U^{t*}_{j2}C_0^h).
$$

As before, we define
\begin{eqnarray*}
f_m^b&\equiv&\sum_{\beta=1}^3f_m^{b\beta},
\end{eqnarray*}
as shown in Fig.\ref{gb2tw}, where
$$
f_1^{b1}=\frac{3\alpha_seg_sV_{tb}}{\sqrt{2}\pi\sin\theta_W}
\sum_i^2\sum_j^2U^t_{j1}U^{b*}_{i1}U^b_{i2}(M_{\tilde{g}}U^{t*}_{j1}D_{00}^a+
m_tU^{t*}_{j2}D_{002}^a),$$$$
f_2^{b1}=-\frac{3\alpha_seg_sV_{tb}}{\sqrt{2}\pi\sin\theta_W}
\sum_i^2\sum_j^2U^t_{j1}U^{b*}_{i1}U^b_{i2}(M_{\tilde{g}}U^{t*}_{j1}D_{13}^a+
m_tU^{t*}_{j2}D_{123}^a),$$$$
f_3^{b1}=\frac{3\alpha_seg_sV_{tb}}{\sqrt{2}\pi\sin\theta_W}
\sum_i^2\sum_j^2U^t_{j1}U^{b*}_{i1}[M_{\tilde{g}}U^b_{i2}U^{t*}_{j1}(D_1^a+D_{11}^a+D_{12}^a)+
m_tU^b_{i2}U^{t*}_{j2}(D_{112}^a+D_{12}^a+D_{122}^a)],$$$$
f_4^{b1}=\frac{3\alpha_seg_sV_{tb}}{\sqrt{2}\pi\sin\theta_W}
\sum_i^2\sum_j^2U^t_{j1}U^{b*}_{i1}[M_{\tilde{g}}U^b_{i2}U^{t*}_{j1}(D_{12}^a+D_2^a+D_{22}^a)+
m_tU^b_{i2}U^{t*}_{j2}(D_{122}^a+D_{22}^a+D_{222}^a)],$$$$
f_5^{b1}=-\frac{3\alpha_seg_sV_{tb}}{\sqrt{2}\pi\sin\theta_W}
\sum_i^2\sum_j^2U^t_{j1}U^{b*}_{i1}U^b_{i2}(M_{\tilde{g}}U^{t*}_{j1}D_{23}^a+
m_tU^{t*}_{j2}D_{223}^a),$$$$
f_7^{b1}=-\frac{3\alpha_seg_sV_{tb}}{2\sqrt{2}\pi\sin\theta_W}
\sum_i^2\sum_j^2U^t_{j1}U^{b*}_{i1}U^b_{i1}U^{t*}_{j1}D_{00}^a,$$$$
f_8^{b1}=\frac{3\alpha_seg_sV_{tb}}{\sqrt{2}\pi\sin\theta_W}
\sum_i^2\sum_j^2U^t_{j1}U^{b*}_{i1}U^b_{i1}U^{t*}_{j1}D_{001}^a,$$$$
f_9^{b1}=\frac{3\alpha_seg_sV_{tb}}{\sqrt{2}\pi\sin\theta_W}
\sum_i^2\sum_j^2U^t_{j1}U^{b*}_{i1}U^b_{i1}U^{t*}_{j1}D_{002}^a,$$$$
f_{10}^{b1}=-\frac{3\alpha_seg_sV_{tb}}{\sqrt{2}\pi\sin\theta_W}
\sum_i^2\sum_j^2U^t_{j1}U^{b*}_{i1}U^b_{i1}U^{t*}_{j1}(D_{002}^a+D_{003}^a),$$$$
f_{11}^{b1}=-\frac{3\alpha_seg_sV_{tb}}{\sqrt{2}\pi\sin\theta_W}
\sum_i^2\sum_j^2U^t_{j1}U^{b*}_{i1}U^b_{i1}U^{t*}_{j1}(D_{112}^a+
D_{113}^a+D_{12}^a+D_{122}^a+D_{13}^a+D_{123}^a),$$$$
f_{12}^{b1}=-\frac{3\alpha_seg_sV_{tb}}{\sqrt{2}\pi\sin\theta_W}
\sum_i^2\sum_j^2U^t_{j1}U^{b*}_{i1}U^b_{i1}U^{t*}_{j1}(D_{12}^a+D_{122}^a+D_2^a+
2D_{22}^a+D_{222}^a +D_{223}^a+D_{23}^a+D_{123}^a),$$$$
f_{13}^{b1}=\frac{3\alpha_seg_sV_{tb}}{\sqrt{2}\pi\sin\theta_W}
\sum_i^2\sum_j^2U^t_{j1}U^{b*}_{i1}
U^b_{i1}U^{t*}_{j1}(D_{133}^a+D_{123}^a),$$$$
f_{14}^{b1}=\frac{3\alpha_seg_sV_{tb}}{\sqrt{2}\pi\sin\theta_W}
\sum_i^2\sum_j^2U^t_{j1}U^{b*}_{i1}
U^b_{i1}U^{t*}_{j1}(D_{223}^a+D_{23}^a+D_{233}^a),$$$$
f_{15}^{b1}=-\frac{3\alpha_seg_sV_{tb}}{2\sqrt{2}\pi\sin\theta_W}
\sum_i^2\sum_j^2U^t_{j1}U^{b*}_{i1}U^b_{i1}U^{t*}_{j1}(2D_{003}^a-C_2^i-2D_{00}^a+m_t^2D_{23}^a-
tD_{23}^a),$$$$
f_{16}^{b1}=-\frac{3\alpha_seg_sV_{tb}}{2\sqrt{2}\pi\sin\theta_W}
\sum_i^2\sum_j^2U^t_{j1}U^{b*}_{i1}U^b_{i1}U^{t*}_{j1}
[(t-m_t^2)(D_{12}^a+D_2^a+D_{22}^a)-C_2^i-2(D_{00}^a+D_{001}^a+
D_{002}^a)],$$$$
f_{18}^{b1}=-\frac{3\alpha_seg_sV_{tb}}{2\sqrt{2}\pi\sin\theta_W}
\sum_i^2\sum_j^2U^t_{j1}U^{b*}_{i1}M_{\tilde{g}}U^b_{i2}U^{t*}_{j1}D_3^a,$$$$
f_{19}^{b1}=\frac{3\alpha_seg_sV_{tb}}{2\sqrt{2}\pi\sin\theta_W}
\sum_i^2\sum_j^2U^t_{j1}U^{b*}_{i1}[M_{\tilde{g}}U^b_{i2}U^{t*}_{j1}(D_0^a+D_1^a+D_2^a)+
m_tU^b_{i2}U^{t*}_{j2}(D_{12}^a+D_2^a+D_{22}^a-D_{23}^a)],$$$$
f_{20}^{b1}=-\frac{3\alpha_seg_sV_{tb}}{\sqrt{2}\pi\sin\theta_W}
\sum_i^2\sum_j^2U^t_{j1}U^{b*}_{i1}U^b_{i1}(m_tU^{t*}_{j1}D_{123}^a+
M_{\tilde{g}}U^{t*}_{j2}D_{13}^a),$$$$
f_{21}^{b1}=\frac{3\alpha_seg_sV_{tb}}{\sqrt{2}\pi\sin\theta_W}
\sum_i^2\sum_j^2U^t_{j1}U^{b*}_{i1}[m_tU^b_{i1}U^{t*}_{j1}(D_{112}^a+D_{12}^a+D_{122}^a)+
M_{\tilde{g}}U^b_{i1}U^{t*}_{j2}(D_1^a+D_{11}^a+D_{12}^a)],$$$$
f_{22}^{b1}=\frac{3\alpha_seg_sV_{tb}}{\sqrt{2}\pi\sin\theta_W}
\sum_i^2\sum_j^2U^t_{j1}U^{b*}_{i1}U^b_{i1}(m_tU^{t*}_{j1}D_{002}^a+
M_{\tilde{g}}D_{00}^a),$$$$
f_{23}^{b1}=\frac{3\alpha_seg_sV_{tb}}{\sqrt{2}\pi\sin\theta_W}
\sum_i^2\sum_j^2U^t_{j1}U^{b*}_{i1}[m_tU^b_{i1}U^{t*}_{j1}(D_{122}^a+D_{22}^a+D_{222}^a)+
M_{\tilde{g}}U^b_{i1}U^{t*}_{j2}(D_{12}^a+D_2^a+D_{22}^a)],$$$$
f_{24}^{b1}=-\frac{3\alpha_seg_sV_{tb}}{\sqrt{2}\pi\sin\theta_W}
\sum_i^2\sum_j^2U^t_{j1}U^{b*}_{i1}U^b_{i1}
(m_tU^{t*}_{j1}D_{223}^a+M_{\tilde{g}}U^{t*}_{j2}D_{23}^a),$$$$
f_{25}^{b1}=-\frac{3\alpha_seg_sV_{tb}}{2\sqrt{2}\pi\sin\theta_W}
\sum_i^2\sum_j^2U^t_{j1}U^{b*}_{i1}U^b_{i2}U^{t*}_{j2}(m_t^2D_{23}^a
-tD_{23}^a-A_{25}- 2D_{00}^a+ 2D_{003}^a),$$$$
f_{26}^{b1}=-\frac{3\alpha_seg_sV_{tb}}{2\sqrt{2}\pi\sin\theta_W}
\sum_i^2\sum_j^2U^t_{j1}U^{b*}_{i1}U^b_{i2}
U^{t*}_{j2}[(t-m_t^2)(D_{12}^a+D_2^a+D_{22}^a) -C_2^i-2(D_{00}^a+
D_{001}^a+D_{002}^a)],$$$$
f_{27}^{b1}=-\frac{3\alpha_seg_sV_{tb}}{2\sqrt{2}\pi\sin\theta_W}
\sum_i^2\sum_j^2U^t_{j1}U^{b*}_{i1}U^b_{i2}U^{t*}_{j2}D_{00}^a,$$$$
f_{28}^{b1}=-\frac{3\alpha_seg_sV_{tb}}{2\sqrt{2}\pi\sin\theta_W}
\sum_i^2\sum_j^2U^t_{j1}U^{b*}_{i1}U^b_{i1}(m_tU^{t*}_{j1}D_{23}^a+
M_{\tilde{g}}U^{t*}_{j2}D_3^a),$$$$
f_{29}^{b1}=-\frac{3\alpha_seg_sV_{tb}}{2\sqrt{2}\pi\sin\theta_W}
\sum_i^2\sum_j^2U^t_{j1}U^{b*}_{i1}[m_tU^b_{i1}U^{t*}_{j1}(D_{12}^a+D_2^a+D_{22}^a)+
M_{\tilde{g}}U^b_{i1}U^{t*}_{j2}(D_0^a+D_1^a+D_2^a)],$$$$
f_{30}^{b1}=\frac{3\alpha_seg_sV_{tb}}{\sqrt{2}\pi\sin\theta_W}
\sum_i^2\sum_j^2U^t_{j1}U^{b*}_{i1}U^b_{i2}
U^{t*}_{j2}D_{001}^a,$$$$
f_{31}^{b1}=\frac{3\alpha_seg_sV_{tb}}{\sqrt{2}\pi\sin\theta_W}
\sum_i^2\sum_j^2U^t_{j1}U^{b*}_{i1}U^b_{i2}U^{t*}_{j2}D_{002}^a,$$$$
f_{32}^{b1}=-\frac{3\alpha_seg_sV_{tb}}{\sqrt{2}\pi\sin\theta_W}
\sum_i^2\sum_j^2U^t_{j1}U^{b*}_{i1}U^b_{i2}U^{t*}_{j2}(D_{002}^a+D_{003}^a),$$$$
f_{33}^{b1}=-\frac{3\alpha_seg_sV_{tb}}{\sqrt{2}\pi\sin\theta_W}
\sum_i^2\sum_j^2U^t_{j1}U^{b*}_{i1}U^b_{i2}U^{t*}_{j2}(D_{112}^a+
D_{113}^a+D_{12}^a+D_{122}^a+ D_{13}^a+D_{123}^a),$$$$
f_{34}^{b1}=-\frac{3\alpha_seg_sV_{tb}}{\sqrt{2}\pi\sin\theta_W}
\sum_i^2\sum_j^2U^t_{j1}U^{b*}_{i1}U^b_{i2}U^{t*}_{j2}(D_{12}^a+D_{122}^a+D_2^a+2D_{22}^a+
D_{222}^a+ D_{223}^a+D_{23}^a+D_{123}^a),$$$$
f_{35}^{b1}=\frac{3\alpha_seg_sV_{tb}}{\sqrt{2}\pi\sin\theta_W}
\sum_i^2\sum_j^2U^t_{j1}U^{b*}_{i1}U^b_{i2}U^{t*}_{j2}(D_{133}^a+D_{123}^a),$$$$
f_{36}^{b1}=\frac{3\alpha_seg_sV_{tb}}{\sqrt{2}\pi\sin\theta_W}
\sum_i^2\sum_j^2U^t_{j1}U^{b*}_{i1}U^b_{i2}U^{t*}_{j2}(D_{223}^a+D_{23}^a+D_{233}^a),
$$
$$
f_1^{b2}=\frac{\alpha_seg_sV_{tb}}{3\sqrt{2}\pi\sin\theta_W}\sum_i^2\sum_j^2
U^t_{j1}U^{b*}_{i1}(M_{\tilde{g}}U^b_{i2}U^{t*}_{j1}D_{00}^b+
m_tU^b_{i2}U^{t*}_{j2}D_{003}^b),$$$$
f_2^{b2}=-\frac{\alpha_seg_sV_{tb}}{3\sqrt{2}\pi\sin\theta_W}\sum_i^2\sum_j^2
U^t_{j1}U^{b*}_{i1}[M_{\tilde{g}}U^b_{i2}U^{t*}_{j1}(D_1^b+
D_{12}^b+ D_{23}^b+D_3^b+D_{33}^b+D_{13}^b)$$$$
+m_tU^b_{i2}U^{t*}_{j2}(D_{123}^b+D_{13}^b+
D_{233}^b+D_{33}^b+D_{333}^b+D_{133}^b)],$$$$
f_3^{b2}=\frac{\alpha_seg_sV_{tb}}{3\sqrt{2}\pi\sin\theta_W}\sum_i^2\sum_j^2
U^t_{j1}U^{b*}_{i1}[M_{\tilde{g}} U^b_{i2}U^{t*}_{j1}(D_{11}^b+
D_{13}^b) +m_tU^b_{i2}U^{t*}_{j2}(D_{113}^b+ D_{133}^b)],$$$$
f_4^{b2}=-\frac{\alpha_seg_sV_{tb}}{3\sqrt{2}\pi\sin\theta_W}\sum_i^2\sum_j^2
U^t_{j1}U^{b*}_{i1}(M_{\tilde{g}}U^b_{i2}U^{t*}_{j1}D_{13}^b
+m_tU^b_{i2}U^{t*}_{j2}D_{133}^b),$$$$
f_5^{b2}=\frac{\alpha_seg_sV_{tb}}{3\sqrt{2}\pi\sin\theta_W}\sum_i^2\sum_j^2
U^t_{j1}U^{b*}_{i1}[M_{\tilde{g}}U^b_{i2}U^{t*}_{j1}
(D_{23}^b+D_3^b+D_{33}^b)+m_tU^b_{i2}U^{t*}_{j2}(D_{233}^b+
D_{33}^b+D_{333}^b)],$$$$
f_8^{b2}=-\frac{\alpha_seg_sV_{tb}}{3\sqrt{2}\pi\sin\theta_W}\sum_i^2\sum_j^2
U^t_{j1}U^{b*}_{i1}U^b_{i1}U^{t*}_{j1}(D_{001}^b+D_{003}^b),$$$$
f_9^{b2}=\frac{\alpha_seg_sV_{tb}}{3\sqrt{2}\pi\sin\theta_W}\sum_i^2\sum_j^2
U^t_{j1}U^{b*}_{i1}U^b_{i1}U^{t*}_{j1}D_{003}^b,$$$$
f_{10}^{b2}=\frac{\alpha_seg_sV_{tb}}{3\sqrt{2}\pi\sin\theta_W}\sum_i^2\sum_j^2
U^t_{j1}U^{b*}_{i1}U^b_{i1}U^{t*}_{j1}D_{002}^b,$$$$
f_{11}^{b2}=\frac{\alpha_seg_sV_{tb}}{3\sqrt{2}\pi\sin\theta_W}\sum_i^2\sum_j^2
U^t_{j1}U^{b*}_{i1}U^b_{i1}U^{t*}_{j1}(D_{112}^b+D_{123}^b),$$$$
f_{12}^{b2}=-\frac{\alpha_seg_sV_{tb}}{3\sqrt{2}\pi\sin\theta_W}\sum_i^2\sum_j^2
U^t_{j1}U^{b*}_{i1}U^b_{i1}U^{t*}_{j1}D_{123}^b,$$$$
f_{13}^{b2}=-\frac{\alpha_seg_sV_{tb}}{3\sqrt{2}\pi\sin\theta_W}\sum_i^2\sum_j^2
U^t_{j1}U^{b*}_{i1}U^b_{i1}U^{t*}_{j1}(D_{223}^b+D_{23}^b+
D_{233}^b+ D_{12}^b+ D_{122}^b+D_{123}^b),$$$$
f_{14}^{b2}=\frac{\alpha_seg_sV_{tb}}{3\sqrt{2}\pi\sin\theta_W}\sum_i^2\sum_j^2
U^t_{j1}U^{b*}_{i1}U^b_{i1}U^{t*}_{j1}(D_{223}^b+D_{23}^b
+D_{233}^b),$$$$
f_{15}^{b2}=\frac{\alpha_seg_sV_{tb}}{3\sqrt{2}\pi\sin\theta_W}\sum_i^2\sum_j^2
U^t_{j1}U^{b*}_{i1}U^b_{i1}U^{t*}_{j1}(D_{00}^b+D_{002}^b+
D_{003}^b),$$$$
f_{16}^{b2}=-\frac{\alpha_seg_sV_{tb}}{3\sqrt{2}\pi\sin\theta_W}\sum_i^2\sum_j^2
U^t_{j1}U^{b*}_{i1}U^b_{i1}U^{t*}_{j1}D_{001}^b,$$$$
f_{20}^{b2}=-\frac{\alpha_seg_sV_{tb}}{3\sqrt{2}\pi\sin\theta_W}\sum_i^2\sum_j^2
U^t_{j1}U^{b*}_{i1}[m_tU^b_{i1}U^{t*}_{j1}(D_{123}^b+
D_{13}^b+D_{233}^b+D_{33}^b +D_{333}^b+ D_{133}^b)$$$$
+M_{\tilde{g}}U^b_{i1}U^{t*}_{j2}(D_1^b+ D_{12}^b+ D_{23}^b+
D_3^b+D_{33}^b+D_{13}^b)],$$$$
f_{21}^{b2}=\frac{\alpha_seg_sV_{tb}}{3\sqrt{2}\pi\sin\theta_W}\sum_i^2\sum_j^2
U^t_{j1}U^{b*}_{i1}[m_tU^b_{i1}U^{t*}_{j1}(D_{113}^b+ D_{133}^b)+
M_{\tilde{g}}U^b_{i1}U^{t*}_{j2}(D_{11}^b+ D_{13}^b)],$$$$
f_{22}^{b2}=\frac{\alpha_seg_sV_{tb}}{3\sqrt{2}\pi\sin\theta_W}\sum_i^2\sum_j^2
U^t_{j1}U^{b*}_{i1}(m_tU^b_{i1}U^{t*}_{j1}D_{003}^b+
M_{\tilde{g}}U^b_{i1}U^{t*}_{j2}D_{00}^b),$$$$
f_{23}^{b2}=-\frac{\alpha_seg_sV_{tb}}{3\sqrt{2}\pi\sin\theta_W}\sum_i^2\sum_j^2
U^t_{j1}U^{b*}_{i1}(m_tU^b_{i1}U^{t*}_{j1}D_{133}^b+
M_{\tilde{g}}U^b_{i1}U^{t*}_{j2}D_{13}^b),$$$$
f_{24}^{b2}=\frac{\alpha_seg_sV_{tb}}{3\sqrt{2}\pi\sin\theta_W}\sum_i^2\sum_j^2
U^t_{j1}U^{b*}_{i1}[M_{\tilde{g}}U^b_{i1}U^{t*}_{j2}(D_{23}^b+
D_3^b+D_{33}^b) +m_tU^b_{i1}U^{t*}_{j1}(D_{233}^b+D_{33}^b
+D_{333}^b)],$$$$
f_{25}^{b2}=\frac{\alpha_seg_sV_{tb}}{3\sqrt{2}\pi\sin\theta_W}\sum_i^2\sum_j^2
U^t_{j1}U^{b*}_{i1}U^b_{i2}U^{t*}_{j2}(D_{00}^b+D_{002}^b+
D_{003}^b),$$$$
f_{26}^{b2}=-\frac{\alpha_seg_sV_{tb}}{3\sqrt{2}\pi\sin\theta_W}\sum_i^2\sum_j^2
U^t_{j1}U^{b*}_{i1}U^b_{i2}U^{t*}_{j2}D_{001}^b,$$$$
f_{30}^{b2}=-\frac{\alpha_seg_sV_{tb}}{3\sqrt{2}\pi\sin\theta_W}\sum_i^2\sum_j^2
U^t_{j1}U^{b*}_{i1}U^b_{i2}U^{t*}_{j2}(D_{001}^b+D_{003}^b),$$$$
f_{31}^{b2}=\frac{\alpha_seg_sV_{tb}}{3\sqrt{2}\pi\sin\theta_W}\sum_i^2\sum_j^2
U^t_{j1}U^{b*}_{i1}U^b_{i2}U^{t*}_{j2}D_{003}^b,$$$$
f_{32}^{b2}=\frac{\alpha_seg_sV_{tb}}{3\sqrt{2}\pi\sin\theta_W}\sum_i^2\sum_j^2
U^t_{j1}U^{b*}_{i1}U^b_{i2}U^{t*}_{j2}D_{002}^b,$$$$
f_{33}^{b2}=\frac{\alpha_seg_sV_{tb}}{3\sqrt{2}\pi\sin\theta_W}\sum_i^2\sum_j^2
U^t_{j1}U^{b*}_{i1}U^b_{i2}U^{t*}_{j2}(D_{112}^b+ D_{123}^b),$$$$
f_{34}^{b2}=-\frac{\alpha_seg_sV_{tb}}{3\sqrt{2}\pi\sin\theta_W}\sum_i^2\sum_j^2
U^t_{j1}U^{b*}_{i1}U^b_{i2}U^{t*}_{j2}D_{123}^b,$$$$
f_{35}^{b2}=-\frac{\alpha_seg_sV_{tb}}{3\sqrt{2}\pi\sin\theta_W}\sum_i^2\sum_j^2
U^t_{j1}U^{b*}_{i1}U^b_{i2}U^{t*}_{j2}(D_{12}^b+ D_{122}^b
+D_{223}^b+D_{23}^b+D_{233}^b + D_{123}^b),$$$$
f_{36}^{b2}=\frac{\alpha_seg_sV_{tb}}{3\sqrt{2}\pi\sin\theta_W}\sum_i^2\sum_j^2
U^t_{j1}U^{b*}_{i1}U^b_{i2}U^{t*}_{j2}(D_{223}^b+D_{23}^b+D_{233}^b),
$$
$$
f_1^{b3}=-\frac{\alpha_seg_sV_{tb}}{3\sqrt{2}\pi\sin\theta_W}
\sum_i^2\sum_j^2U^t_{i1}U^{b*}_{j1}[m_tU^b_{j2}U^{t*}_{i2}(D_{00}^c+D_{001}^c
+D_{003}^c) -M_{\tilde{g}}U^b_{j2}U^{t*}_{i1}D_{00}^c],$$$$
f_2^{b3}=-\frac{\alpha_seg_sV_{tb}}{3\sqrt{2}\pi\sin\theta_W}
\sum_i^2\sum_j^2U^t_{i1}U^{b*}_{j1}[M_{\tilde{g}}U^b_{j2}U^{t*}_{i1}(D_{23}^c+
D_{13}^c+D_3^c+D_{33}^c)$$$$
-m_tU^b_{j2}U^{t*}_{i2}(D_{113}^c+2D_{133}^c
+D_{123}^c+2D_{13}^c+D_{23}^c+D_{233}^c +D_3^c
+2D_{33}^c+D_{333}^c)],$$$$
f_3^{b3}=-\frac{\alpha_seg_sV_{tb}}{3\sqrt{2}\pi\sin\theta_W}
\sum_i^2\sum_j^2U^t_{i1}U^{b*}_{j1}[M_{\tilde{g}}U^b_{j2}U^{t*}_{i1}D_{13}^c
-m_tU^b_{j2}U^{t*}_{i2}(D_{113}^c+D_{133}^c +D_{13}^c)],$$$$
f_4^{b3}=-\frac{\alpha_seg_sV_{tb}}{3\sqrt{2}\pi\sin\theta_W}
\sum_i^2\sum_j^2U^t_{i1}U^{b*}_{j1}[m_tU^b_{j2}U^{t*}_{i2}(D_{11}^c+D_{111}^c+
2D_{113}^c +D_{133}^c+D_{13}^c)$$$$
-M_{\tilde{g}}U^b_{j2}U^{t*}_{i1}(D_{11}^c+D_{13}^c)],$$$$
f_5^{b3}=-\frac{\alpha_seg_sV_{tb}}{3\sqrt{2}\pi\sin\theta_W}
\sum_i^2\sum_j^2U^t_{i1}U^{b*}_{j1}[m_tU^b_{j2}U^{t*}_{i2}(2D_{11}^c
+D_{111}^c + 3D_{113}^c +3D_{133}^c +2D_{123}^c +4D_{13}^c$$$$
+D_{23}^c+D_{233}^c +D_3^c+2D_{33}^c+D_{333}^c
+D_1^c+D_{112}^c+D_{12}^c)$$$$
-M_{\tilde{g}}U^b_{j2}U^{t*}_{i1}(D_1^c+D_{11}^c+D_{12}^c+D_{23}^c+
2D_{13}^c+D_3^c+D_{33}^c)],$$$$
f_8^{b3}=\frac{\alpha_seg_sV_{tb}}{3\sqrt{2}\pi\sin\theta_W}
\sum_i^2\sum_j^2U^t_{i1}U^{b*}_{j1}U^b_{j1}U^{t*}_{i1}D_{003}^c,$$$$
f_9^{b3}=-\frac{\alpha_seg_sV_{tb}}{3\sqrt{2}\pi\sin\theta_W}
\sum_i^2\sum_j^2U^t_{i1}U^{b*}_{j1}U^b_{j1}U^{t*}_{i1}(D_{001}^c-D_{003}^c),$$$$
f_{10}^{b3}=-\frac{\alpha_seg_sV_{tb}}{3\sqrt{2}\pi\sin\theta_W}
\sum_i^2\sum_j^2U^t_{i1}U^{b*}_{j1}U^b_{j1}U^{t*}_{i1}D_{002}^c,$$$$
f_{11}^{b3}=\frac{\alpha_seg_sV_{tb}}{3\sqrt{2}\pi\sin\theta_W}
\sum_i^2\sum_j^2U^t_{i1}U^{b*}_{j1}U^b_{j1}U^{t*}_{i1}D_{123}^c,$$$$
f_{12}^{b3}=-\frac{\alpha_seg_sV_{tb}}{3\sqrt{2}\pi\sin\theta_W}
\sum_i^2\sum_j^2U^t_{i1}U^{b*}_{j1}U^b_{j1}U^{t*}_{i1}(D_{112}^c+D_{123}^c),$$$$
f_{13}^{b3}=\frac{\alpha_seg_sV_{tb}}{3\sqrt{2}\pi\sin\theta_W}
\sum_i^2\sum_j^2U^t_{i1}U^{b*}_{j1}U^b_{j1}U^{t*}_{i1}(D_{123}^c+D_{223}^c+D_{23}^c+D_{233}^c),$$$$
f_{14}^{b3}=-\frac{\alpha_seg_sV_{tb}}{3\sqrt{2}\pi\sin\theta_W}
\sum_i^2\sum_j^2U^t_{i1}U^{b*}_{j1}U^b_{j1}U^{t*}_{i1}(D_{112}^c+D_{12}^c+
D_{122}^c +2D_{123}^c+D_{223}^c+D_{23}^c+D_{233}^c),$$$$
f_{15}^{b3}=-\frac{\alpha_seg_sV_{tb}}{3\sqrt{2}\pi\sin\theta_W}
\sum_i^2\sum_j^2U^t_{i1}U^{b*}_{j1}U^b_{j1}U^{t*}_{i1}(D_{00}^c+
2D_{001}^c+ D_{002}^c+ D_{003}^c),$$$$
f_{20}^{b3}=-\frac{\alpha_seg_sV_{tb}}{3\sqrt{2}\pi\sin\theta_W}
\sum_i^2\sum_j^2U^t_{i1}U^{b*}_{j1}[M_{\tilde{g}}U^b_{j1}U^{t*}_{i2}(D_{13}^c+D_{23}^c
+D_3^c +D_{33}^c)$$$$
-m_tU^b_{j1}U^{t*}_{i1}(D_{113}^c+2D_{13}^c+D_{123}^c+2D_{133}^c+D_{23}^c+D_{233}^c+D_3^c
+2D_{33}^c+D_{333}^c)],$$$$
f_{21}^{b3}=-\frac{\alpha_seg_sV_{tb}}{3\sqrt{2}\pi\sin\theta_W}
\sum_i^2\sum_j^2U^t_{i1}U^{b*}_{j1}[M_{\tilde{g}}U^b_{j1}U^{t*}_{i2}D_{13}^c
-m_tU^b_{j1}U^{t*}_{i1}(D_{113}^c+D_{13}^c+D_{133}^c)],$$$$
f_{22}^{b3}=-\frac{\alpha_seg_sV_{tb}}{3\sqrt{2}\pi\sin\theta_W}
\sum_i^2\sum_j^2U^t_{i1}U^{b*}_{j1}[m_tU^b_{j1}U^{t*}_{i1}(D_{00}^c+D_{001}^c+D_{003}^c)-
M_{\tilde{g}}U^b_{j1}U^{t*}_{i2}D_{00}^c],$$$$
f_{23}^{b3}=-\frac{\alpha_seg_sV_{tb}}{3\sqrt{2}\pi\sin\theta_W}
\sum_i^2\sum_j^2U^t_{i1}U^{b*}_{j1}[m_tU^b_{j1}U^{t*}_{i1}(D_{11}^c+D_{111}^c+2D_{113}^c
+D_{13}^c+D_{133}^c)$$$$
-M_{\tilde{g}}U^b_{j1}U^{t*}_{i2}(D_{11}^c+D_{13}^c)],$$$$
f_{24}^{b3}=-\frac{\alpha_seg_sV_{tb}}{3\sqrt{2}\pi\sin\theta_W}
\sum_i^2\sum_j^2U^t_{i1}U^{b*}_{j1}[m_tU^b_{j1}U^{t*}_{i1}
(D_1^c+2D_{11}^c+D_{112}^c+D_{111}^c+3D_{113}^c
+D_{12}^c+4D_{13}^c$$$$
+2D_{123}^c+3D_{133}^c+D_{23}^c+D_{233}^c+D_3^c
+2D_{33}^c+D_{333}^c)$$$$
-M_{\tilde{g}}U^b_{j1}U^{t*}_{i2}(D_{11}^c+D_{12}^c+D_1^c
+2D_{13}^c+D_{23}^c +D_3^c +D_{33}^c)],$$$$
f_{25}^{b3}=-\frac{\alpha_seg_sV_{tb}}{3\sqrt{2}\pi\sin\theta_W}
\sum_i^2\sum_j^2U^t_{i1}U^{b*}_{j1}U^b_{j2}U^{t*}_{i2}
(D_{00}^c+D_{002}^c+D_{003}^c+D_{001}^c),$$$$
f_{26}^{b3}=-\frac{\alpha_seg_sV_{tb}}{3\sqrt{2}\pi\sin\theta_W}
\sum_i^2\sum_j^2U^t_{i1}U^{b*}_{j1}U^b_{j2}U^{t*}_{i2}D_{001}^c,$$$$
f_{30}^{b3}=\frac{\alpha_seg_sV_{tb}}{3\sqrt{2}\pi\sin\theta_W}
\sum_i^2\sum_j^2U^t_{i1}U^{b*}_{j1}U^b_{j2}U^{t*}_{i2}D_{003}^c,$$$$
f_{31}^{b3}=-\frac{\alpha_seg_sV_{tb}}{3\sqrt{2}\pi\sin\theta_W}
\sum_i^2\sum_j^2U^t_{i1}U^{b*}_{j1}U^b_{j2}U^{t*}_{i2}(D_{001}^c+D_{003}^c),$$$$
f_{32}^{b3}=-\frac{\alpha_seg_sV_{tb}}{3\sqrt{2}\pi\sin\theta_W}
\sum_i^2\sum_j^2U^t_{i1}U^{b*}_{j1}U^b_{j2}U^{t*}_{i2}D_{002}^c,$$$$
f_{33}^{b3}=\frac{\alpha_seg_sV_{tb}}{3\sqrt{2}\pi\sin\theta_W}
\sum_i^2\sum_j^2U^t_{i1}U^{b*}_{j1}U^b_{j2}U^{t*}_{i2}D_{123}^c,$$$$
f_{34}^{b3}=-\frac{\alpha_seg_sV_{tb}}{3\sqrt{2}\pi\sin\theta_W}
\sum_i^2\sum_j^2U^t_{i1}U^{b*}_{j1}U^b_{j2}U^{t*}_{i2}(D_{112}^c+D_{123}^c),$$$$
f_{35}^{b3}=\frac{\alpha_seg_sV_{tb}}{3\sqrt{2}\pi\sin\theta_W}
\sum_i^2\sum_j^2U^t_{i1}U^{b*}_{j1}
U^b_{j2}U^{t*}_{i2}(D_{223}^c+D_{23}^c+D_{233}^c+D_{123}^c),$$$$
f_{36}^{b3}=-\frac{\alpha_seg_sV_{tb}}{3\sqrt{2}\pi\sin\theta_W}
\sum_i^2\sum_j^2U^t_{i1}U^{b*}_{j1}U^b_{j2}U^{t*}_{i2}(D_{223}^c+D_{23}^c+D_{233}^c+D_{112}^c
+D_{12}^c+D_{122}^c+2D_{123}^c).
$$

\newpage

\newpage
\begin{thebibliography}{99}
\bibitem{topquark} the Report of the "1999 CERN Workshop on SM
physics (and more) at the LHC", hep-ph/0003033.
\bibitem{wagner} Wolfgang Wagner, Rep.Prog.Phys. $\mathbf{68}$,
(2005).
\bibitem{topquark123} A.P.Heinson, A.S.Belyaev and
E.E.Boos, Phys. Rev. $\mathbf{D56}$, 3114 (1997) hep-ph/9612424.
\bibitem{topquark129} D.O.Carlson and C.P.Yuan,
Phys. Lett. $\mathbf{B306}$, 386(1993).
\bibitem{topquark130} G.Mahlon and S.Parke, Phys. Rev. $\mathbf{D55}$,
7249 (1997) hep-ph/9611367.
\bibitem{topquark131} G.Mahlon and S.Parke, hep-ph/9912458.
\bibitem{topquark132} D.O.Carlson, E.Malkawi and C.P.Yuan,
Phys. Lett. $\mathbf{B337}$,145 (1994) hep-ph/9405277; A.Datta and
X.Zhang, Phys. Rev. $\mathbf{D55}$, 2530 (1997) hep-ph/9611247.
\bibitem{topquark133} T.Tait and C.P.Yuan, hep-ph/9710372.
\bibitem{topquark134} K.Hikasa, K.Whisnant, J.M.Yang and B.Young,
Phys. Rev. $\mathbf{D58}$, 114003 (1998) hep-ph/9806401.
\bibitem{topquark135} E.Boos, L.Dudko and T.Ohl,
Eur. Phys. J. $\mathbf{C11}$, 473 (1999) hep-ph/9903215.
\bibitem{topquark142} T. Han, M. Hosch, K. Whisnant, B. Young and
X. Zhang, Phys. Rev. $\mathbf{D58}$, 073008 (1998) hep-ph/9806486.
\bibitem{liu1} J. J. Liu, C. S. Li, L. L. Yang and L. G. Jin,
Nucl.\ Phys.\ B {\bf 705}, 3 (2005) hep-ph/0404099.
\bibitem{liu2} J. J. Liu, C. S. Li, L. L. Yang and L. G. Jin,
Mod.\ Phys.\ Lett.\ A {\bf 19}, 317 (2004).
\bibitem{topquark136} D.Atwood, S.Bar-Shalom, G.Eilam and A.Soni,
Phys. Rev. $\mathbf{D54}$, 5412 (1996) hep-ph/9605345.
\bibitem{topquark137} E.H.Simmons,
Phys. Rev. $\mathbf{D55}$, 5494 (1997) hep-ph/9612402.
\bibitem{topquark138} C.S.Li, R.J.Oakes and J.M.Yang,
Phys. Rev. $\mathbf{D55}$, 1672 (1997) hep-ph/9608460.
\bibitem{topquark139} C.S.Li, R.J.Oakes and J.M.Yang,
Phys. Rev. $\mathbf{D55}$, 5780 (1997) hep-ph/9611455; C.S.Li,
R.J.Oakes, J.M.Yang and H.Y.Zhou, Phys. Rev. $\mathbf{D57}$,
2009(1998) hep-ph/9706412.
\bibitem{topquark140} S.Bar-Shalom, D.Atwood and A.Soni,
Phys. Rev. $\mathbf{D57}$,1495 (1998) hep-ph/9708357.
\bibitem{topquark120} S.S.Willenbrock and D.A.Dicus,
Phys. Rev. $\mathbf{D34}$, 155 (1986); C.P.Yuan, Phys. Rev.
$\mathbf{D41}$, 42 (1990); R.K.Ellis and S.Parke, Phys. Rev.
$\mathbf{D46}$, 3785 (1992).
\bibitem{topquark121} S.Cortese and R.Petronzio,
Phys. Lett. $\mathbf{B253}$, 494 (1991).
\bibitem{topquark122} T.Stelzer and S.Willenbrock,
Phys. Lett. $\mathbf{B357}$, 125 (1995) hep-ph/9505433.
\bibitem{topquark124} T.M.Tait, Phys. Rev. $\mathbf{D61}$,034001 (2000)
hep-ph/9909352.
\bibitem{topquark149} D.Green et al., "A Study of Single Top at
CMS," CMS Note 1999/048, unpublished.
\bibitem{topquark152} B. Gonz$\acute{a}$lez Pi$\tilde{\text{n}}$eiro,
D.O'Neil, R.Brock and M.Lefebvre, "Measuring $V_{tb}$ and
Polarization via Boson Gluon Fusion at the LHC", ATLAS Note:
ATL-COM-PHYS-99-027, unpublished.
\bibitem{topquark30} ATLAS Collab., "ATLAS Detector and Physics
Performance Technical Design Report", CERN LHCC 99-14/15 (1999).
\bibitem{topquark125} T.Stelzer, Z.Sullivan and S.Willenbrock,
Phys. Rev. $\mathbf{D56}$, 5919 (1997) hep-ph/9705398.
\bibitem{topquark126} M.C.Smith and S.Willenbrock,
Phys. Rev. $\mathbf{D54}$, 6696 (1996) hep-ph/9604223.
\bibitem{myself2} Shouhua Zhu, KA-TP-27-2001, hep-ph/0109269.
\bibitem{softgluon} N. Kidonakis, hep-ph/0609287.
\bibitem{myself1} M. Beccaria, G. Macorini, F. M. Renard and
C. Verzegnassi, hep-ph/0605108.
\bibitem{plusgbew} M.~Beccaria, G.~Macorini, F.~M.~Renard and C.~Verzegnassi,
``Associated t W production at LHC: A complete calculation of
electroweak supersymmetric effects at one loop,'' Phys.\ Rev.\ D
{\bf 73}, 093001 (2006) [arXiv:hep-ph/0601175].
\bibitem{li} C.S.Li, R.J.Oakes, J.M.Yang and H.Y.Zhou,
Phys. Rev. $\mathbf{D57}$, 2009 (1998).
\bibitem{li11} A.Sirlin, Phys. Rev. D $\mathbf{22}$, 971 (1980);
W.J.Marciano and A.Sirlin, ibid. $\mathbf{22}$, 2695 (1980);
A.Sirlin and W.J.Marciano, Nucl. Phys. $\mathbf{B189}$, 442 (1981);
K.I.Aoki et al., Prog. Theor. Phys. Suppl. $\mathbf{73}$, 1 (1982).
\bibitem{myself6} J. Collins, F.Wilczek, and A. Zee,
Phys.Rev.$\mathbf{D 18}$, 242 (1978); W.J.Marciano, Phys. Rev.
$\mathbf{D 29}$, 580 (1984); P. Nason, S. Dawson, and R.K. Ellis,
Nucl. Phys. $\mathbf{B 327}$, 49 (1989); Nucl. Phys. $\mathbf{B
335}$, 260 (1989) (E)
\bibitem{zhao14} S. G. Gorishny et al., Mod. Phys. Lett. A $\mathbf{5}$,
2703(1990); Phys. Rev. D $\mathbf{43}$, 1633(1991); A. Djouadi et
al., Z. Phys. C $\mathbf{70}$ 427(1996); Comput. Phys. Commun.
$\mathbf{108}$, 56(1998); M. Spira, Fortschr. Phys. $\mathbf{46}$,
203(1998).
\bibitem{pdg} Review of Particle Physics(2004).
\bibitem{zhao16} J.Pumplin et al., J.High Energy Phys. 07(2002) 012.
\bibitem{zhao19} M.Drees and S.P.Martin, hep-ph/9504324.
\bibitem{zhao20} A.Djouadi et al., hep-ph/0211331.
\bibitem{myself3} V. M. Abazov et al., Fermilab-Pub-06/06-077-E,
hep-ex/0604029.
\bibitem{myself4} A. Abulencia et al., hep-ex/0512072.
\bibitem{myself5} A. Denner, Fortschr. Phys.$\mathbf{41}$, 307
(1993).
\end{thebibliography}
\end{document}